\documentclass[titlepage,12pt]{utarticle}
\usepackage{amssymb, amsmath, amsopn, amsthm,amscd,amsfonts}
\usepackage{color}
\usepackage{latexsym}
\usepackage{ifthen}
\usepackage{epsfig}
\usepackage{graphicx}
\usepackage{graphics}
\usepackage{mathtools}
\usepackage{bm,longtable,enumerate}
\numberwithin{equation}{section}
\newcommand{\ket}[1]{\lvert #1\rangle}

\newcommand{\tr}{{\rm tr}}

\newcommand{\wt}{\widetilde}
\newcommand{\wh}{\widehat}
\newcommand{\ol}{\overline}
\newcommand{\del}{\partial}

\newcommand{\nn}{\nonumber}

\newcommand{\AD}[1]{$\ol{\mbox{D~\,}}\!\!\!$#1}

\newcommand{\cA}{{\cal A}}

\newcommand{\cF}{{\cal F}}

\newcommand{\bR}{\mathbb{R}}
\newcommand{\eqn}{\begin{eqnarray}}
\newcommand{\eqnx}{\end{eqnarray}}

\def\beq{\begin{equation}}
\def\eeq{\end{equation}}
\def\beqa{\begin{eqnarray}}
\def\eeqa{\end{eqnarray}}
\def\tr{\mathop{\rm tr}\nolimits}

\def\matt[#1,#2,#3,#4]{\left(%
\begin{array}{cc} #1 & #2 \\ #3 & #4 \end{array} \right)}

\def\v2#1{\vv2[#1]}
\def\vv2[#1,#2]{\left(%
{#1 \atop #2}\right)}

\def\ol{\overline}

\def\nn{\nonumber}

\def\beq{\begin{equation}}
\def\eeq{\end{equation}}
\def\beqa{\begin{eqnarray}}
\def\eeqa{\end{eqnarray}}

\begin{document}
\preprint{DIAS-STP-11-10 \\
          DCPT-11/53\\}
\title{{\bf Generalized baryon form factors and \vskip 0.0cm proton structure functions \vskip 0.3cm in the Sakai-Sugimoto model}}

\author{{\sc C.~A.~Ballon Bayona}
 \address{
  	Centre for Particle Theory, \\
         University of Durham, \\
	 Science Laboratories, South Road,\\
         Durham DH1 3LE, United Kingdom.\\[0.1cm]
         {\tt c.a.m.ballonbayona@durham.ac.uk}
         } 
\\ {\sc Henrique Boschi-Filho $^{\rm b}$, Nelson R.~F.~Braga}  
 \address{
 Instituto de F{\'i}sica, \\
 Universidade Federal do Rio de Janeiro,\\
         Caixa Postal 68528,\\
         21941-972 Rio de Janeiro, RJ, Brasil.\\[0.1cm]
         {\tt boschi@if.ufrj.br},\\
         {\tt braga@if.ufrj.br},\\
         {\tt mtorres@if.ufrj.br}
         	        } ,
{\sc Matthias Ihl}
 \address{
        School of Theoretical Physics, \\
        Dublin Institute for Advanced Studies, \\
        10 Burlington Rd, \\
        Dublin 4, Ireland. \\[0.1cm]
        {\tt msihl@stp.dias.ie}
  }
\\ {\sc Marcus A.~C.~Torres} $^{\rm b}$ 
}

\Abstract{We investigate the production of positive parity baryon resonances in proton electromagnetic scattering within the Sakai-Sugimoto model. 
The latter is a string model for the non-perturbative regime of large $N_c$ QCD. Using holographic techniques
 we calculate the generalized Dirac and Pauli form factors that describe resonance production.
 We use these results to estimate the contribution of resonance production to the proton structure functions. Interestingly, we find an approximate Callan-Gross relation for  
the structure functions in a regime of intermediate values of the Bjorken variable.}

\maketitle
\tableofcontents
\newpage

\section{Introduction}

In the  regime of low momentum transfer ($\sqrt{q^2}$ lower than a few $GeVs$), non-perturbative effects become relevant in hadronic scattering. 
Lattice QCD (the numerical approach of substituting the continuous spacetime by a lattice of points) is quite difficult to use here because of 
the real time dependence in scattering processes. Effective models can be very useful but they have a low predictability because they 
may depend on many parameters.

\medskip

In recent years, gauge/string duality has evolved into an invaluable tool to study many strongly coupled phenomena in particle and condensed 
matter physics. In particular, it has brought a new insight into the strongly coupled regime of Yang-Mills theories. 
A good example is the successful prediction of the shear viscosity for strongly coupled theories whose gravity dual 
involves a black hole in Anti-de-Sitter space \cite{hep-th/0405231}. This prediction is in accordance with the small value 
observed in the quark-gluon plasma phase observed in heavy ion collisions at RHIC and LHC. 
The study of 10-d string models (top-down approach) and 5-d effective models (bottom-up approach) dual to 4-d QCD-like theories is known as AdS/QCD (see \cite{Peeters:2007ab,Erdmenger:2007cm,Gubser:2009md,arXiv:1101.0618}
for a review of the top-down approach,  \cite{hep-th/0610135,arXiv:0908.0312,arXiv:1001.1978} for the bottom-up approach and \cite{arXiv:1006.5461} for an hybrid approach). 
AdS/QCD is a powerful approach to the strong coupling regime of QCD because the string models depend on very few parameters (the string length, string coupling and number of 
colors and flavours of the dual theory). On top of that, general properties of  hadron phenomenology (like vector meson dominance) seem to be universal in this approach in the 
sense that they do not depend on the particular AdS/QCD model. There is one string model, though, that has a field content very similar to large $N_c$ QCD 
in the regime of large distances and massless quarks. This is the Sakai-Sugimoto model \cite{Sakai:2004cn,Sakai:2005yt} that realizes confinement and chiral symmetry breaking.

\medskip

In this paper we investigate the production of baryon resonances with positive parity in proton electromagnetic scattering within the Sakai-Sugimoto model. 
First we define the current matrix element that describe  the transition from a proton to a positive parity baryonic resonance. We derive the general expansion
of the current matrix element in terms of scalars that we define as generalized Dirac and Pauli form factors which include as a particular case the well known 
elastic Dirac and Pauli form factors. Using holographic techniques we obtain a relation between the generalized Dirac and Pauli form factors 
and the couplings between baryons and vector mesons. This result can be interpreted as a realization of vector meson dominance in electromagnetic scattering 
of baryons.

\medskip 

We also estimate in this paper the contribution of resonance production to the proton structure functions. 
The latter are Lorentz invariant scalars defined in Deep Inelastic Scatering (DIS) which is the inclusive scattering of a proton by a virtual photon (emitted by a lepton). 
The proton structure functions have a nice interpretation in terms of parton distribution functions that describe probability densities of finding partons (valence quarks, gluons and 
quark-antiquark pairs) with a fraction of the longitudinal momentum of the proton. A calculation of these quantities at low momentum transfer proves difficult in non-perturbative QCD  and usually relies heavily on some input from 
experiments or simulations. It should be noted that in the Sakai-Sugimoto model we can only make reliable predictions about  scattering processes with low momentum transfers (
in this paper $q^2 \le 5 ({\text GeV})^2$). This regime  is very far from the Bjorken limit of DIS ($q^2 \to \infty$) which is well described by perturbative QCD. Moreover, it is beyond the scope
 of this paper to study  the full inclusive DIS process with arbitrary final states. Therefore we limit ourselves to the contribution coming from single final state baryons with the same 
spin and isospin as the proton but different masses. 

\medskip 

Electromagnetic form factors have been obtained previously using bottom-up (phenomenological) models and top-down string models. 
The meson form factors have been calculated in \cite{Grigoryan:2007vg,Grigoryan:2007my,arXiv:0707.3859} (bottom-up) 
and \cite{Hong:2003jm,RodriguezGomez:2008zp,BallonBayona:2009ar,Bayona:2010bg} (top-down). Baryon form factors have been obtained in \cite{arXiv:0903.4818,arXiv:0904.3731} (bottom-up)
and \cite{Hong:2007ay,Hashimoto:2008zw,Kim:2008pw,arXiv:0904.3710} (top-down). Deep Inelastic Scattering in AdS/QCD was first investigated by Polchinski and Strassler 
in a bottom-up model for the case of scalar glueballs and baryon-like fermions \cite{Polchinski:2002jw}. 
Further development of DIS in bottom-up and top-down models include the large $x$ regime  \cite{BallonBayona:2007qr,BallonBayona:2008zi,Pire:2008zf,BallonBayona:2010ae} 
as well as the small $x$ regime where Pomeron exchange dominates \cite{Brower:2006ea,Hatta:2007he,BallonBayona:2007rs,arXiv:0804.1562,Cornalba:2009ax,Brower:2010wf}.
DIS structure functions have also been calculated for strongly coupled plasmas \cite{Hatta:2007cs,Bayona:2009qe,Iancu:2009py,Bu:2011my}.

\medskip 

The discussion in section \ref{sec:EMscattering} is independent of a specific model realization and contains novel results that apply very generally to non-elastic electromagnetic 
scattering for baryons, namely a detailed derivation of electromagnetic current matrix elements and structure functions.
In section \ref{sec:SakaiSugimotoBaryons} we briefly review the Sakai-Sugimoto model and the description of holographic baryons in this context. Section \ref{sec:GFF} contains the derivation of 
the electromagnetic current and the generalized Dirac and Pauli form factors. Numerical results for the wave functions, masses, couplings and form factors are presented for the lowest excited states of spin $1/2$ and positive parity.
In section \ref{sec:PSF} we present our numerical estimate on the proton structure functions and the associated Callan-Gross relation. 

\section{Baryon resonances in proton electromagnetic scattering}\label{sec:EMscattering}

Massless QCD with $N_f$ flavours enjoys chiral symmetry ($U(N_f)_L \times U(N_f)_R$) at high energies. At low energies chiral symmetry is spontaneously broken and the residual symmetry is vectorial 
corresponding to the group $U(N_f)_V$. In the case of two flavours, the vectorial current can be written as $J_V^{\mu,a}$ where $a=(0,1,2,3)$. 
The electromagnetic current ${\cal J}^\mu$ can be obtained as a combination of the isoscalar current $J_V^{\mu,0}$ and the isovector current $J_V^{\mu,3}$ :  
\beqa
{\cal J}^\mu = \frac{1}{N_c} J_V^{\mu,0} + J_V^{\mu,3} \equiv \sum_{a=0}^3 c_a J_V^{\mu,a} \,. \label{emcurrent}
\eeqa
In this paper we are interested in  the production of baryon resonances with positive parity in electromagnetic scattering of a proton. Namely, we study the transition of a proton (denoted by $B$) with spin $1/2$, isospin $1/2$  and momentum $p$ 
to a baryonic resonance  (denoted by $B_{ X}$) with the same spin and isospin as the proton but momentum $p_{ X}$ and different mass. This transition is characterized by evaluating the matrix elements of electromagnetic currents between the baryonic initial and final states. 
We denote the spin and isospin projections of the initial (final)  baryon state as $s$ ($s_{ X}$) and $I_3$ ($I_3^{  X}$) respectively. 

\subsection{The generalized Dirac and Pauli form factors}

Evaluating the isoscalar and isovector current operators in the baryonic states we obtain current matrix elements that can be decomposed as 
\beqa
\langle p_{  X} , B_{  X} , s_{  X} , I_3^{  X}  \vert J_V^{\mu,a} (0) \vert p , B, s , I_3 \rangle &=& \frac{i}{2 (2 \pi)^3}  (\tau^a)_{I_3^{  X} I_3}
 \bar u (p_{  X}  , s_{  X}) \Big [  \gamma^\mu F^{D,a}_{B B_{  X}}(q^2)  \cr
&+& \kappa_B  \sigma^{\mu \nu} q_\nu  F^{P,a}_{B B_{  X}} (q^2) \, +\, i q^\mu   F^{Q,a}_{B B_{  X}}(q^2) \Big ] u (p , s)  \, \label{CurrentMatrix},
\eeqa
where
\beqa
q^\mu &=& (p_{  X} - p)^\mu \quad , \quad \kappa_B = \frac{1}{m_B + m_{B_{  X}} } \, , \cr
(\tau^0)_{I_3^{  X} I_3} &=& \delta_{I_3^{  X} I_3} \quad , \quad 
(\tau^a)_{I_3^{  X} I_3} = (\sigma^a)_{I_3^{  X} I_3} \quad a=(1,2,3) \,,
\eeqa
and $\sigma^a$ are the Pauli matrices. In (\ref{CurrentMatrix}) we have used a generalization of the Gordon decomposition identity
\beqa
{\bar u}_{B_X}(p') \gamma^{\mu} u_B(p) = {\bar u}_{B_X}(p') \left[ - \frac{p^{\prime \mu} + p^{\mu}}{m_B + m_{B_X}}+\frac{ i \sigma^{\mu \nu}\left(p'_{\nu}-p_{\nu}\right)}{m_B + m_{B_X}}\right] u_B(p) \,. 
\eeqa
The scalars $F^{D,a}_{B B_{  X}}(q^2)$,$F^{P,a}_{B B_{  X}}(q^2)$ in (\ref{CurrentMatrix}) are the Dirac and Pauli form factors while $F^{Q,a}_{B B_{  X}}(q^2)$ is required by current conservation
\beqa
0 &=& q_\mu \langle  J_V^{\mu,a} (0) \rangle \sim 
 \bar u (p_{  X}  , s_{  X}) \Big [  (p_{  X} - p)_\mu \gamma^\mu F^{D,a}_{B B_{  X}}(q^2)  \,+\,  i q^2   F^{Q,a}_{B B_{  X}}(q^2) \Big ] u (p , s) \cr
&=& \bar u (p_{  X}  , s_{  X}) \Big [
i ( m_{B_{  X}} - m_B ) F^{D,a}_{B B_{  X}}(q^2)  \,+\, i  q^2   F^{Q,a}_{B B_{  X}}(q^2) \Big ] u (p , s) \, ,
\eeqa
so that
\beqa
 F^{Q,a}_{B B_{  X}}(q^2) = - \frac{1}{q^2} ( m_{B_{  X}} - m_B ) F^{D,a}_{B B_{  X}}(q^2) \,.
\eeqa
The current matrix element now takes a transverse form 
\beqa
\langle p_{  X} , B_{  X} , s_{  X}  \vert J_V^{\mu,a} (0) \vert p , B, s \rangle &=& \frac{i}{2 (2 \pi)^3}  (\tau^a)_{I_3^{  X} I_3}
\left ( \eta^{\mu \nu}  - \frac{ q^\mu q^\nu}{q^2} \right )
 \bar u (p_{  X}  , s_{  X}) \Big [  \gamma_\nu F^{D,a}_{B B_{  X}}(q^2)  \cr
&+& \kappa_B  \sigma_{\nu \lambda} q^\lambda  F^{P,a}_{B B_{  X}} (q^2)  \Big ] u (p , s)  \, . \label{matrixelement}
\eeqa
We have chosen the signature $\eta_{\mu \nu}={\rm diag}(-,+,+,+)$. The relativistic Dirac spinors  that represent the initial and final states can be written as  
\beqa
u(p,s) = \frac{1}{\sqrt{2E}} \left( \begin{array}{c} f \, \chi_s (\vec{p}) \\ \frac{\vec{p} \cdot \vec{\sigma}}{f} \, \chi_s (\vec{p}) \end{array} \right) \quad , \quad
u(p_{  X},s_{  X}) = \frac{1}{\sqrt{2 E_{  X}}} \left( \begin{array}{c} f_{  X} \, \chi_{s_{  X}} (\vec{p}_{ X})  \\ \frac{\vec{p}_{  X} \cdot \vec{\sigma}}{f_{  X}} 
\, \chi_{s_{  X}} (\vec{p}_{ X}) \end{array} \right) \, , 
\eeqa
where $f = \sqrt{E + m_B}$ and $f_{  X} = \sqrt{E_{  X} + m_{B_{  X}}}$. The two-component spinors $\chi_s (\vec{p})$ and $\chi_{s_{  X}} (\vec{p}_{ X})$ are defined as eigenstates of the helicity operators 
\beqa
\frac{\vec{p} \cdot \vec{\sigma}}{|\vec{p}|} \chi_s (\vec{p})  = s \, \chi_s (\vec{p}) \quad , \quad 
\frac{\vec{p}_{ X} \cdot \vec{\sigma}}{|\vec{p}_{ X} |} \chi_{s_{  X}} (\vec{p}_{ X}) = s_{ X} \, \chi_{s_{  X}} (\vec{p}_{ X}) \, .
\eeqa
The convention for gamma matrices is 
\beqa
\gamma^0  = - i \left( \begin{array}{cc} 1 & 0  \\ 0 & - 1 \end{array} \right) \quad, \quad \gamma^i  = - i \left( \begin{array}{cc} 0 & \sigma^i  \\ -\sigma^i  & 0 \end{array} \right)
\quad , \quad
\sigma^{\mu \nu} = \frac{i}{2} \left [ \gamma^\mu , \gamma^\nu \right ] \, .
\eeqa
The baryon states are normalized according to their charges as 
\beqa
 \langle p_{  X} , B_{  X} , s_{  X}, I_3^{  X} \vert  Q_V^a \vert p , B, s, I_3  \rangle  
&=& \langle p_{  X} , B_{  X} , s_{  X}, I_3^{  X}   \vert J_V^{0,a} (0) \vert p , B, s, I_3  \rangle (2 \pi)^3 \delta^3 (\vec{q})  \cr
&=& \frac12 (\tau^a)_{I_3^{  X} I_3} \delta^3 (\vec{q}) \delta_{s_{  X} s} \delta_{B B_{  X}}  F^{D,a}_{B B}(0) \,.
\label{chargenorm}
\eeqa
In the case where the initial state is a proton and the final state has the same isospin polarization ($I_3^{  X}=I_3 =1/2$) 
the isoscalar and isovector charges are $N_c/2$  and $1/2$ respectively. The elastic Dirac form factors are fixed at $q^2 =0$ as 
\beqa
F^{D,0}_{BB}(0) = N_c \quad , \quad F^{D,3}_{BB}(0)=1 \, ,
\eeqa 
so  we find from (\ref{chargenorm}) 
\beqa
\langle p_{  X}, B_{ X} , s_{ X} ,  1/2 \vert  p , B , s ,  1/2 \rangle 
&=&  \delta^3 (\vec {p}_{  X} - \vec{p}) \delta_{s_{ X} s}
\delta_{B_{ X} B} \, .
\eeqa
The following spinor relations are very useful  
\beqa
\bar u (p_{  X},s_{  X}) \gamma^0 u(p,s)  &=&  - \frac{i}{2 \sqrt{E E_{ X}}} \chi_{s_{  X}}^\dag (\vec{p}_{ X}) \Big [ f f_{ X}    
\,+\,  \frac{1 }{f f_{ X}}  \vec{p}_{ X} \cdot \vec{\sigma} \vec{p} \cdot \vec{\sigma} \Big ] \chi_s (\vec{p})  \cr
&=& - \frac{i}{2 \sqrt{E E_{ X}}} \left ( \frac{f}{f_{ X}} \right )
\left [  f_X^2 + \frac{s_{ X} s | \vec{p}_{ X} | |\vec{p}  | }{ f^2 } \right ] \chi_{s_{  X}}^\dag (\vec{p}_{ X}) \chi_s (\vec{p}) \, , \cr
&& \label{spinrel1}
\eeqa
\beqa
\bar u (p_{  X},s_{  X}) \gamma^i u(p,s)  &=&  - \frac{i}{2 \sqrt{E E_{ X} } } \chi_{s_{  X}}^\dag (\vec{p}_{ X}) \Big \{ \frac{f_{ X}}{f}  \sigma^i \vec{p} \cdot \vec{\sigma} 
\,+\, \frac{f}{f_{ X}} \vec{p}_{  X} \cdot \vec{\sigma} \sigma^i \Big \} \chi_s (\vec{p})  \cr
&=&  - \frac{i}{2 \sqrt{E E_{ X} } } \left ( \frac{f}{f_{ X}} \right ) 
\left [ \frac{f_X^2 }{f^2} s | \vec{p} | + s_{ X} | \vec{p}_{ X} | \right ]
\, \chi_{s_{  X}}^\dag (\vec{p}_{ X})  \sigma^i \chi_s(\vec{p}) \, , \cr
&& \label{spinrel2}
\eeqa
\beqa
\bar u (p_{  X},s_{  X}) \sigma^{0i} q_i u(p,s) &=& - \frac{i}{2 \sqrt{E E_{ X}}}  \left [ \frac{f_{ X} }{f} q^i p^j - \frac{f}{f_{ X}} (p + q)^i q^j \right ]
\, \chi_{s_{  X}}^\dag (\vec{p}_{ X})  \sigma_i \sigma_j \chi_s(\vec{p})   \cr
&=&  - \frac{i}{2 \sqrt{E E_{ X}}} \left ( \frac{f}{f_{ X}} \right ) 
\left [ \frac{f_X^2 }{ f^2} s | \vec{p} | - s_{ X} | \vec{p}_{ X} | \right ]\, q_i \, \chi_{s_{  X}}^\dag (\vec{p}_{ X}) 
\sigma^i \chi_s(\vec{p})  \, , \cr
&& \label{spinrel3}
\eeqa
\beqa
\bar u (p_{  X},s_{  X}) \sigma^{ij} q_j u(p,s) &=& - \frac{1}{2 \sqrt{E E_{ X}}} \epsilon^{ijk} q_j
\chi_{s_{  X}}^\dag (\vec{p}_{ X}) \Big [ f_{ X} f \sigma_k 
\,-\, \frac{1}{f_{ X}f} (p + q)^a p^b \sigma_a \sigma_k \sigma_b \Big ] \chi_s(\vec{p}) \cr
&=&  - \frac{1}{2 \sqrt{E E_{ X}}} \left ( \frac{f}{f_{ X}} \right ) \epsilon^{ijk} q_j 
\left [  f_X^2 - \frac{s_{ X} s | \vec{p}_{ X} | |\vec{p}  | }{f^2} \right ] 
\chi_{s_{  X}}^\dag (\vec{p}_{ X}) \sigma_k \chi_s(\vec{p}) \,. \cr
&& \label{spinrel4}
\eeqa
{\bf The Breit frame.} 
\medskip
The Breit frame is characterized by the condition $E_{  X}=E$ which means that the photon has zero energy ($q_0=0$). As shown in the appendix, 
choosing the photon in the z axis, the the photon and baryon momenta in the Breit frame take the form 
\beqa
q^\mu &=& (0 , 0 , 0 , q ) \quad , \quad p^\mu = (E , 0 , 0 , p_3 )  \quad, \quad p^\mu_{ X} = p^\mu + q^\mu \, ,  \cr
p_3 &=& - \frac{q}{2x} \quad , \quad E = \sqrt{m_B^2 + p_3^2} \, .
\eeqa
Using eq. (\ref{matrixelement}) and the spinor relations (\ref{spinrel1}) - (\ref{spinrel4}) we can work out the components of the current matrix elements. In the Breit frame the current matrix elements simplify to
\beqa
 \langle p_{  X} , B_{  X} , s_{  X}, I_3^{  X}  \vert J_V^{0,a} (0) \vert p , B, s, I_3  \rangle
&=&  \frac{1}{2 (2 \pi)^3}  (\tau^a)_{I_3^{  X} I_3} \, \chi_{s_{  X}}^\dag (\vec{p}_{ X}) \chi_s (\vec{p}) \cr
&\times& \Big  [ \alpha F^{D,a}_{B B_{  X}}(q^2) -  \beta \, q^2  \kappa_B F^{P,a}_{B B_{  X}}(q^2) \Big ] \, , \label{CurrentBreit0}
\eeqa
\beqa
\langle p_{  X} , B_{  X} , s_{  X}, I_3^{  X}  \vert J_V^{i,a} (0) \vert p , B, s, I_3 \rangle
&=&  - \frac{i}{2 (2 \pi)^3} (\tau^a)_{I_3^{  X} I_3} \epsilon^{ijk} q_j \, \chi_{s_{  X}}^\dag (\vec{p}_{ X}) \sigma_k \chi_s (\vec{p}) \cr 
&\times&  \Big [ \beta F^{D,a}_{B B_{  X}}(q^2) + \alpha \kappa_B F^{P,a}_{B B_{  X}}(q^2) \Big ] \, ,  \label{CurrentBreiti}
\eeqa
where 
\beqa
\alpha &=& \left (\frac{1}{2E} \right ) \left ( \frac{\sqrt{E + m_B}}{\sqrt{E + m_{B_{ X}}}}\right ) \left [ E + m_{B_{ X}} + (E - m_B) (1-2x) \right ] \, , \cr
\beta &=&  \left (\frac{1}{2E} \right ) \left ( \frac{\sqrt{E + m_B}}{\sqrt{E + m_{B_{ X}}}}\right ) \left ( \frac{1}{2x} \right ) \left [ \frac{E + m_{B_{ X}} }{E + m_B} + 2x - 1 \right ] \, , \cr
E &=& \sqrt{m_B^2 + \frac{q^2}{4 x^2}} \, . \label{alphabeta}
\eeqa
To get (\ref{CurrentBreit0}) and (\ref{CurrentBreiti}), the following identity proved useful,
\beqa
s \chi_{s_{  X}}^\dag (\vec{p}_{ X}) \sigma^i \chi_s (\vec{p}) = 
- \delta^{i3} \chi_{s_{  X}}^\dag (\vec{p}_{ X}) \chi_s (\vec{p}) 
- i \epsilon^{i3k} \chi_{s_{  X}}^\dag (\vec{p}_{ X}) \sigma^k \chi_s(\vec{p}) \, , \label{helident}
\eeqa
which is valid  in the Breit frame only. \\
Note that in the elastic case $m_{B_{  X}} = m_B$, $f_{  X}=f$ and $\kappa_B = 1/(2m_B)$ so we obtain
\beqa
 \langle p_{  X} , B_{  X} , s_{  X}, I_3^{  X}  \vert J_V^{0,a} (0) \vert p , B, s, I_3 \rangle
&=&  \frac{1}{2 (2 \pi)^3}  (\tau^a)_{I_3^{  X} I_3} \left ( \frac{m_B}{E} \right ) \, \chi_{s_{  X}}^\dag (\vec{p}_{ X}) \chi_s (\vec{p})
G^{E,a}_{B}(q^2) \, , \cr
\langle p_{  X} , B_{  X} , s_{  X}, I_3^{  X}  \vert J_V^{i,a} (0) \vert p , B, s, I_3  \rangle
&=&  - \frac{1}{2 (2 \pi)^3}  (\tau^a)_{I_3^{  X} I_3} \left ( \frac{i}{2E} \right )
\epsilon^{ijk} q_j \, \chi_{s_{  X}}^\dag (\vec{p}_{ X}) \sigma_k \chi_s (\vec{p}) 
G^{M,a}_{B}(q^2) \, , \cr
&&
\eeqa
where 
\beqa
 G^{E,a}_{B}(q^2) = F^{D,a}_{B}(q^2) -  \frac{q^2}{4 m_B^2} F^{P,a}_{B}(q^2) \quad , \quad 
G^{M,a}_{B}(q^2) = F^{D,a}_{B}(q^2) + F^{P,a}_{B}(q^2) \, ,
\eeqa
are the elastic electric and magnetic form factors also known as the Sachs form factors.

\subsection{Deep Inelastic Scattering and the proton structure functions}

Deep inelastic scattering (DIS) is a primary tool to investigate the internal structure of hadrons. DIS refers to the scattering process of a lepton on a hadron. The lepton interacts with a hadron 
of momentum $p^{\mu}$ via a virtual photon of momentum $q^{\mu}$ (cf.~figure 1). The final hadronic state is denoted by $X$ and momentum $p^{\mu}_X$. Such a process is commonly parametrized by two dynamical variables, namely the Bjorken parameter $x = -\frac{q^2}{2 p \cdot q}$
and the photon virtuality $q^2$ (for a review of DIS, see e.g., \cite{Manohar:1992tz}). The standard limit in DIS corresponds to the Bjorken limit of large $q^2$ and fixed $x$. In this paper we are interested in the regime of small $q^2$ where non-perturbative contributions are relevant. 

The DIS differential cross section is determined by the hadronic tensor,
\begin{equation}\label{eq:Wmunu}
W^{\mu\nu} \, = \, \frac{1}{8 \pi} \sum_s \int d^4 x \, e^{iq\cdot x} \langle p , s \vert \, \Big[ {\cal J}^\mu (x) , {\cal J}^\nu (0) \Big] 
\, \vert p , s \rangle,
\end{equation}
where ${\cal J}^\mu(x)$ is the electromagnetic current. Inserting the sum of the final states $X$ we can write the hadronic tensor as 
\beqa
W^{\mu \nu} = \frac{1}{8 \pi} \sum_s \sum_X (2 \pi)^4 \delta^4( p + q - p_X)  
\langle p, s  \vert {\cal J}^\mu (0) \vert X \rangle  \langle X \vert {\cal J}^\nu(0) \vert p, s \rangle \, .
\eeqa
The structure functions  $F_1 (x,q^2)$ and $F_2 (x,q^2)$ are Lorentz invariant scalars that appear in the decomposition of the hadronic tensor :  
\begin{equation} \label{eq:strfun}
W^{\mu\nu} \, = \, F_1 (x,q^2)  \Big( \eta^{\mu\nu} \,-\, \frac{q^\mu q^\nu}{q^2} \, \Big) 
\,+\,\frac{2x}{q^2} F_2 (x,q^2)  \Big( p^\mu \,+ \, \frac{q^\mu}{2x} \, \Big) 
\Big( p^\nu \,+ \, \frac{q^\nu}{2x} \, \Big). 
\end{equation}

\begin{figure}\label{fig:DISfig}
\begin{center}
\setlength{\unitlength}{0.1in}
\vskip 3.cm
\begin{picture}(0,0)(15,0)
\rm
\thicklines
\put(1,14.5){$\ell$}
\put(3,15){\line(2,-1){7}}
\put(3,15){\vector(2,-1){4}}
\put(18,14.5){$\ell$}
\put(17,15){\line(-2,-1){7}}
\put(10,11.5){\vector(2,1){4.2}}
\put(9.5,8){$q$}
\bezier{300}(10,11.5)(10.2,10.7)(11,10.5)
\bezier{300}(11,10.5)(11.8,10.3)(12,9.5)
\bezier{300}(12,9.5)(12.2,8.7)(13,8.5)
\bezier{300}(13,8.5)(13.8,8.3)(14,7.5)
\put(0,-2){$p$}
\put(3,0){\line(2,1){10.5}}
\put(3,0){\vector(2,1){6}}
\put(16,6){\circle{5}}
\put(27,-2){$X$}
\put(18.5,5.5){\line(3,-1){8}}
\put(18.3,5){\line(2,-1){8}}
\put(18,4.5){\line(3,-2){7.5}}
\put(17.5,3.8){\line(1,-1){6}}
\end{picture}
\vskip 1.cm
\parbox{4.1 in}{\caption{Exemplary diagram for a deep inelastic scattering process. A lepton $\ell$ exchanges a virtual photon with a hadron of momentum $p$.}}
\end{center}
\end{figure}
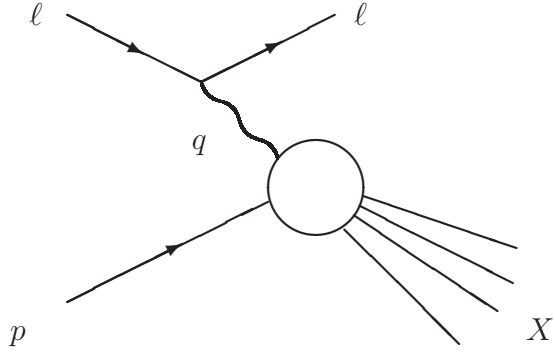
\vskip .5cm
\subsubsection{The contribution from resonance production }

First of all we need to transform the spinors and baryon states of the previous section as 
\beqa
u (p , s)  \to \frac{1}{\sqrt{2 E}} \, u (p,s) \quad , \quad \vert  p , B , s \rangle \to \frac{1}{\sqrt{2 E} (2 \pi)^{3/2}}  \vert  p , B , s \rangle \,, \label{spinortransform} 
\eeqa
in order to get the standard relativistic normalizations 
\beqa
\bar u (p , s) u (p , s) = 2 m_B \quad , \quad \langle p_{  X}, B_{ X} , S_{ X} \vert  p , B , s \rangle = 2 \sqrt{E E_{ X}} (2 \pi)^3  
 \delta^3 (\vec {p}_{  X} - \vec{p}) \delta_{s_{ X} s} \,.
\eeqa
Using (\ref{emcurrent}), (\ref{matrixelement}) and (\ref{spinortransform}) we obtain for $I_3=I_3^{ X}=1/2$,
\beqa
\langle p_{  X} , B_{  X} , s_{  X}  \vert {\cal J}^\mu (0) \vert p , B, s \rangle &=& i  
\left ( \eta^{\mu \nu}  - \frac{ q^\mu q^\nu}{q^2} \right )
 \bar u (p_{  X}  , s_{  X}) \Big [  \gamma_\nu F^{D}_{B B_{  X}}(q^2)  \cr
&&  \quad \quad + \kappa_B  \sigma_{\nu \lambda} q^\lambda  F^{P}_{B B_{  X}} (q^2)  \Big ] u (p , s)  \, , \label{matrixelement2}
\eeqa
where 
\beqa
F^{D}_{B B_{  X}}(q^2) = \frac12 \sum_a c_a F^{D,a}_{B B_{  X}}(q^2) \quad , \quad 
F^{P}_{B B_{  X}}(q^2) = \frac12 \sum_a c_a F^{P,a}_{B B_{  X}}(q^2) \,,
\eeqa
are the Dirac and Pauli electromagnetic form factors. \\
The baryonic tensor for a spin $1/2$ baryon in the case where one particle is produced in the final state can be written as 
\beqa
W^{\mu \nu} &\,=\,& \frac{1}{8\pi} \sum_{s,s_{ X} } \sum_{m_{B_X}} \int \frac{d^4 p_X}{(2 \pi)^3} \theta(p_X^0) \delta( p_X^2 + m_{B_X}^2) \cr 
&& \quad \times (2 \pi)^4 \delta^4(p + q - p_X) \langle p , B , s | {\cal J}^{\mu}(0) | p_X , B_X , s_{ X} \rangle  \langle p_X , B_X  , s_{ X} | {\cal J}^{\nu}(0) | p , B \rangle \cr 
&& \cr
&\,=\,& \frac14  \sum_{s,s_{ X}} \sum_{m_{B_X}} \delta \left [ ( p+ q)^2 + m_{B_X}^2  \right ]
 \langle p , B ,s | {\cal J}^{\mu}(0) | p_X , B_X , s_{ X} \rangle  \langle p_X , B_X , s_{ X} | {\cal J}^{\nu}(0) | p , B , s \rangle \, . \cr 
&& \label{hadronictensor}
\eeqa
Substituting (\ref{matrixelement2}) into (\ref{hadronictensor}) we obtain 
\beqa
W^{\mu \nu}&\,=\,& - \frac{1}{4}  \sum_{m_{B_X}} \delta \left [ ( p+ q)^2 + m_{B_X}^2  \right ]  
\left ( \eta^{\mu \rho} - \frac{q^\mu q^\rho}{q^2} \right ) \left ( \eta^{\nu \sigma } - \frac{q^\nu q^\sigma}{q^2} \right) \cr
&&\quad \times \Big [ F^{D}_{B B_X}(q^2) F^{D}_{B B_X}(q^2) {\cal A}_{\rho \sigma} 
+ F^{P}_{B B_X}(q^2) F^{P}_{B B_X}(q^2) {\cal B}_{\rho \sigma}  \cr 
&& \qquad \,+\, F^{P}_{B B_X}(q^2) F^{D}_{B B_X}(q^2) {\cal C}_{\rho \sigma} 
+ F^{D}_{B B_X}(q^2) F^{P}_{B B_X}(q^2) {\cal D}_{\rho \sigma} \Big ] \label{hadronictensor2}\, ,
\eeqa
where

\beqa
{\cal A}_{\rho \sigma}  &\,=\,& - p^\tau (p + q)^{\bar \tau} \tr (\gamma_\tau \gamma_\rho \gamma_{\bar \tau} \gamma_\sigma) 
+ m_B m_{B_X} \tr (\gamma_\rho \gamma_\sigma) \cr 
&\,=\,& 4 \Big \{ \left [ m_B m_{B_{ X}} + p \cdot ( p + q ) \right ] \eta_{\rho \sigma} 
- 2 p_\rho p_\sigma - p_\rho q_\sigma - p_\sigma q_\rho \Big \} \, ,  \cr
&\,=\,& 4 \Big \{ \left [ m_B m_{B_{ X}} + p^2 - \frac{q^2}{2x} \right ] \eta_{\rho \sigma} 
- 2 p_\rho p_\sigma - p_\rho q_\sigma - p_\sigma q_\rho \Big \} \, ,
\label{calA}
\eeqa
\beqa
{\cal B}_{\rho \sigma}  &\,=\,&  - \kappa_B^2 q^\lambda q^{\bar \lambda} 
\left [ m_B m_{B_X} \tr (\sigma_{\lambda \rho} \sigma_{\bar \lambda \sigma} ) 
- p^\tau (p + q)^{\bar \tau} \tr (  \sigma_{\lambda \rho} \gamma_{\bar \tau} \sigma_{\bar \lambda \sigma} \gamma_\tau ) \right ]  \cr 
&\,=\,& 4 \kappa_B^2 q^2 \Big \{ \left [ - m_B m_{B_{ X}} + p^2 + \frac{q^2}{2x} \left (1 - \frac{1}{x} \right )\right ] \eta_{\rho \sigma} - 2 p_\rho p_\sigma \cr 
&-& \frac{1}{x} ( q_\rho p_\sigma + q_\sigma p_\rho) 
+ \left [ m_B m_{B_{ X}} - p^2 - \frac{q^2}{2x} \right ]\frac{q_\rho q_\sigma}{q^2} \Big \} \, , 
\label{calB}
\eeqa
\beqa
{\cal C}_{\rho \sigma}  &\,=\,& -  i \kappa_B q^\lambda
 \left [ m_B ( p + q)^\tau \tr ( \sigma_{\lambda \rho} \gamma_\tau \gamma_\sigma ) 
+ m_{B_X} p^\tau \tr (\sigma_{\lambda \rho} \gamma_\sigma \gamma_\tau )  \right ]  \cr 
&\,=\,& 4 \kappa_B \Big \{ \left [ - m_B (p + q) \cdot q + m_{B_{ X}} p \cdot q \right ] \eta_{\rho \sigma}
+ \left [ m_B ( p + q)_\rho - m_{B_{ X}} p_\sigma \right ] q_\sigma \Big \} \cr 
&\,=\,& 4 \kappa_B q^2 \Big \{ -  \left [ m_B + \frac{1}{2x} ( m_{B_{ X}} - m_B ) \right ] \eta_{\rho \sigma} 
+ \left [ m_B q_\rho + (m_B - m_{B_{ X}} ) p_\rho \right ] \frac{q_\sigma}{q^2} \Big \},
\label{calC}
\eeqa
and 
\beqa
{\cal D}_{\rho \sigma}  &\,=\,&  i \kappa_B q^\lambda \left [ m_B (p + q)^\tau \tr ( \sigma_{\lambda \sigma} \gamma_\rho \gamma_\tau )
+  m_{B_X} p^\tau \tr (\sigma_{\lambda \sigma} \gamma_\tau \gamma_\rho ) \right ]  \cr 
&\,=\,&  {\cal C}_{\sigma \rho} \,, 
\label{calD}
\eeqa
and we used the sum over spin formula 
\beqa
\sum_{s} u(p,s) \bar u(p,s) = -i \gamma^\mu p_\mu + m_B \, , 
\eeqa
and the gamma trace identities 
\beqa
\tr (\gamma_\mu \gamma_\nu) &=& + 4 \eta_{\mu \nu} \, ,\cr 
\tr (\gamma_\mu \gamma_\nu \gamma_\rho \gamma_\sigma ) &=& + 4 ( \eta_{\mu \nu} \eta_{\rho \sigma} - \eta_{\mu \rho} \eta_{\nu \sigma} + \eta_{\mu \sigma} \eta_{\nu \rho})
 \equiv 4 \tilde \eta_{\mu \nu \rho \sigma} \, , \cr  
\tr (\gamma_\mu \gamma_\nu \gamma_\rho \gamma_\sigma \gamma_\lambda \gamma_\tau ) &=& + 4 \Big [ \eta_{\mu \nu} \tilde \eta_{\rho \sigma \lambda \tau} 
-   \eta_{\mu \rho} \tilde \eta_{\nu \sigma \lambda \tau} +   \eta_{\mu \sigma} \tilde \eta_{\nu \rho \lambda \tau}   
-   \eta_{\mu \lambda} \tilde \eta_{\nu \rho \sigma \tau} +   \eta_{\mu \tau} \tilde \eta_{\nu \rho \sigma \lambda } \Big ] \, , \cr
\tr (\sigma_{\mu \nu} \gamma_\rho \gamma_\sigma) &=& - \tr ( \sigma_{\mu \nu} \gamma_\sigma \gamma_\rho ) 
= 4 i ( - \eta_{\mu \rho} \eta_{\nu \sigma} + \eta_{\mu \sigma} \eta_{\nu \rho}) \, , \cr 
\tr (\sigma_{\mu \nu} \sigma_{\rho \sigma}) &=& - 4 ( - \eta_{\mu \rho}  \eta_{\nu \sigma} + \eta_{\mu \sigma} \eta_{\nu \rho}) \, , \cr
\tr (\sigma_{\mu \nu} \gamma_\rho \sigma_{\sigma \lambda} \sigma_\tau ) &=& 
- 4 \Big [ - \eta_{\mu \rho} ( \eta_{\nu \sigma} \eta_{\lambda \tau} - \eta_{\nu \lambda} \eta_{\sigma \tau}) 
+ \eta_{\mu \sigma} \tilde \eta_{\nu \rho \lambda \tau} - \eta_{\mu \lambda} \tilde \eta_{\nu \rho \sigma \tau} \cr
&& \quad + \eta_{\mu \tau} ( - \eta_{\nu \sigma} \eta_{\rho \lambda} + \eta_{\nu \lambda} \eta_{\rho \sigma} ) \Big ] \, .
\eeqa
Note that the terms with $q_\rho$ or $q_\sigma$ will vanish when contracting with the transverse tensors.
Using (\ref{calA}),(\ref{calB}),(\ref{calC}),(\ref{calD}) in (\ref{hadronictensor2}) we obtain  the baryonic tensor
\beqa
W^{\mu \nu} \,=\, \left ( \eta^{\mu \nu} - \frac{q^\mu q^\nu }{q^2} \right ) F_1 (q^2 ,x) + \left ( p^\mu + \frac{q^\mu}{2x} \right) \left ( p^\nu + \frac{q^\nu}{2x} \right) \frac{2x}{q^2} F_2(q^2,x) \, , 
\eeqa
in terms of the structure functions
\beqa
F_1 (q^2 ,x) &\,=\,&   \sum_{m_{B_X}} \delta \left [ ( p+ q)^2 + m_{B_X}^2  \right ]   
\Big \{  \left [-  m_B ( m_{B_X} - m_B) + \frac{q^2}{2x} \right ]  (F^D_{B B_X} (q^2 , x))^2 \cr
&+&  \left [ m_B (m_{B_X} + m_B)  + \frac{q^2}{2x} \left( \frac{1}{x} - 1 \right) \right ] \kappa_B^2 \, q^2 (F^P_{B B_X}(q^2))^2  \cr
&+& 2  \left [ m_B + \frac{1}{2x} ( m_{B_{ X}} - m_B)  \right ] \kappa_B \, q^2 F^P_{B B_X}(q^2) F^D_{B B_X}(q^2) \Big \}  \, , 
\eeqa
and
\beqa
F_2 (q^2 ,x) &\,=\,&   \left (\frac{q^2}{x} \right )  \sum_{m_{B_X}} \delta \left [ ( p+ q)^2 + m_{B_X}^2  \right ]   \cr 
&&\quad \times \Big [ (F^D_{B B_X}(q^2))^2  + \kappa_B^2 q^2 (F^P_{B B_X}(q^2))^2 \Big ] \,.
\eeqa
Interestingly, we can rewrite $F_1(q^2 , x)$ as a binomial squared, i.e.,
\beqa
F_1 (q^2 ,x) &\,=\,&    \sum_{m_{B_X}} \delta \left [ ( p+ q)^2 + m_{B_X}^2  \right ]   
\zeta^2 \Big [  F^D_{B B_X} (q^2 , x) + F^P_{B B_X} (q^2 , x) \Big ]^2  \, , 
\eeqa
where 
\beqa
\zeta \equiv q \, ( m_B + m_{B_{ X}} )^{-1/2} \left [ m_B + \frac{1}{2x} ( m_{B_{ X}} - m_B) \right ]^{1/2} 
= \frac{1}{\sqrt{2}} \left [ (m_{B_{ X}} - m_B)^2 + q^2 \right ]^{1/2}\,.
\eeqa
Note that in the elastic case $\zeta = q /\sqrt{2}$ and $\kappa_B = 1/(2 m_B)$ so that the structure functions reduce to 
\beqa
F_1 (q^2 ,x) &\,=\,& \left (\frac{q^2}{2} \right) \delta \left ( q^2 + 2 p \cdot q \right ) 
\left [ F^D_{B B}(q^2) + F^P_{B B}(q^2) \right ]^2    \, , \cr
F_2 (q^2 ,x) &\,=\,& q^2 \delta \left ( q^2 + 2 p \cdot q \right )
\left [ (F^D_{B B}(q^2))^2 + \frac{q^2}{4 m_B^2} (F^P_{B B}(q^2))^2 \right ]  \,.
\eeqa

\subsubsection{The helicity amplitudes}\label{helicityamp}

It is interesting to compare the result that we have obtained for the structure functions with the standard result in terms of the 
helicity amplitudes \cite{Carlson:1998gf}
\beqa
F_1 (q^2 ,x) &\,=\,&    \sum_{m_{B_X}} \delta \left [ ( p+ q)^2 + m_{B_X}^2  \right ] m_B^2 (G^+_{B B_X}(q^2))^2   \cr 
F_2 (q^2 ,x) &\,=\,&    \sum_{m_{B_X}} \delta \left [ ( p+ q)^2 + m_{B_X}^2  \right ] \left (\frac{q^2}{2x} \right )  
\left ( 1 + \frac{q^2}{4 m_B^2 x^2} \right )^{-1} \cr 
&\times& \left [ (G^+_{B B_X}(q^2))^2  +  2 (G^0_{B B_X}(q^2))^2  \right ] \, . \label{helicities1}
\eeqa
The helicity amplitudes $G^+_{B B_X}(q^2)$ and $G^0_{B B_X}(q^2)$ describe transitions when the initial state is a proton and the final state is a baryon 
of the same spin but different helicities. The amplitude $G_+$ ($G_0$) corresponds to a final baryon with helicity $1/2$ ($-1/2$) and 
spin polarization $-1/2$ ($-1/2$). \\
Comparing (\ref{helicities1}) with our results for the structure functions we get the interesting relations 
\beqa
(G^+_{B B_X}(q^2))^2 &=&  \frac{\zeta^2}{m_B^2} ( F^D_{B B_X}(q^2) + F^P_{B B_X}(q^2) )^2 \, , \cr
(G^0_{B B_X}(q^2))^2 &=&  \left ( \frac{q^2}{2} \right ) \frac{\zeta^2}{m_B^2} 
\left ( \frac{m_{B_X} + m_B}{q^2} F^D_{B B_X}(q^2)  - \frac{1}{m_{B_X} + m_B} F^P_{B B_X}(q^2) \right )^2
\eeqa

\section{Holographic baryons in the Sakai-Sugimoto  model}\label{sec:SakaiSugimotoBaryons}

As pointed out in the introduction, the Sakai-Sugimoto model provides new insight into the problem 
of hadronic scattering in the non-perturbative regime. Moreover, baryons have been successfully incorporated 
into this model by several groups \cite{Hata:2007mb,Hong:2007ay,Seki:2008mu,Hashimoto:2008zw}. 
This development was inspired by the Skyrme model \cite{Skyrme:1962vh} and Witten's original proposal 
of baryon vertices \cite{Witten:1998xy}. In this section we briefly review the Sakai-Sugimoto model and 
describe the construction of holographic baryons.  

\subsection{Review of the model}

The Sakai-Sugimoto model is based on a configuration of $N_c$ D4 branes and $N_f$ D8-\AD8 branes in the limit of large $N_c$ with fixed $N_f$. 
This limit allows a supergravity description and can be interpreted as the quenching limit in QCD. 
The Sakai-Sugimoto model is the first string model that realizes confinement and chiral symmetry breaking. Below we describe this model 
in some detail. 

\medskip 

Consider a set of $N_c$ coincident D4-branes with a compact spatial direction in type IIA supergravity \cite{hep-th/9803131}. The D4-branes generate a background 
with the following metric, dilaton and four-form:
\beqa
ds^2 &=&  \frac{u^{ 3/2}}{R^{ 3/2}} \left [ \eta_{\mu \nu} dx^\mu dx^\nu +  f(u) d\tau^2  \right ]
+  \frac{R^{ 3/2}}{u^{ 3/2}}  \frac{d u^2}{f(u)} + R^{ 3/2} u^{ 1/2} d \Omega_4^2   \, , \cr
f(u) &=& 1 - \frac{u_{\ast}^3}{u^3} \quad , \quad  e^\phi = g_s   \frac{u^{3/4}}{R^{3/4}} \quad , \quad F_4 = \frac{(2 \pi l_s)^3 N_c}{V_{ S^4}} \epsilon_4 \,  ,
\eeqa
where $u_\ast$ is the tip of the cigar geometry generated by the D4 branes and $R = (\pi g_s N_c)^{1/3} \sqrt{\alpha'}$. The $\tau$ coordinate is compact and in order to avoid conical singularities the $\tau$ period is fixed as
\beqa
\delta \tau = \frac{4 \pi}{3} \frac{R^{3/2}}{u_{\ast}^{1/2}}  \, .
\eeqa 
As a consequence, we get a 4-d effective mass scale 
\beqa
M_\ast = \frac{2 \pi}{\delta \tau} = \frac32 \frac{u_\ast^{1/2}}{R^{3/2}} \, .
\eeqa
The $\tau$ compactification is introduced as a mechanism of supersymmetry breaking and confinement. 
Imposing anti-periodic conditions for the fermionic states we get at low energies a four-dimensional 
non-supersymmetric strongly coupled $U(N_c)$ theory at large $N_c$ with 't Hooft constant  given by 
\beq
\lambda = g_{YM}^2 N_c = (2 \pi M_\ast) \,  g_s N_c \, l_s \,.
\eeq
It is convenient to define a pair of dimensionless coordinates $y$ and $z$ defined by the relations
\begin{equation}
u = u_{\ast} \left (1 + y^2 + z^2 \right )^{1/3} \equiv u_{\ast}
K_{y,z}^{1/3} \quad , \quad \tau = \frac{\delta \tau}{2 \pi} \arctan
\left (\frac{z}{y} \right) \,.
\end{equation} 
In terms of these coordinates the metric takes the form
\beqa
{\rm d}s^2 &=& u_{\ast}^{3/2} R^{-3/2}
K_{y,z}^{1/2} \, \eta_{\mu \nu} {\rm d}x^\mu {\rm d}x^\nu \,+\, \frac49 R^{3/2}
u_{\ast}^{1/2} \frac{K_{y,z}^{-5/6}}{y^2 + z^2} \Big [ (z^2 + y^2
K_{y,z}^{1/3}){\rm d}z^2 \cr
&+& (y^2 + z^2 K_{y,z}^{1/3} ) {\rm d}y^2 + 2 y z (1
- K_{y,z}^{1/3}) {\rm d}y {\rm d}z \Big ] + R^{3/2} u_{ \ast}^{1/2} K_{y,z}^{1/6}
{\rm d}\Omega_4^2 \,.  
\eeqa

\medskip 

Now consider $N_f$ coincident D8-\AD8 probe branes living in the background generated by the $N_c$ D4-branes. The probe approximation is guaranteed by the condition $N_f \ll N_c$.
The $N_f$ D8 branes introduce quark degrees of freedom as fundamental strings extending from the D4 branes to the D8 branes. 
The dynamics of the D8 and \AD8 branes is dictated  by the DBI action. It turns out that the solution to the DBI equations smoothly merges the D8 and \AD8 branes 
in the infrared region (small $u$). This is a geometrical realization of chiral symmetry breaking  $U(N_f) \times U(N_f) \to U(N_f)$. In the simplest case 
the solution is just $y=0$ (antipodal solution) and the induced D8-\AD8 metric takes the form,
\beqa
 {\rm d}s_{D8} = u_{\ast}^{3/2} R^{-3/2} K_z^{1/2} \, \eta_{\mu \nu} {\rm d}x^\mu
{\rm d}x^\nu \,+\, \frac49 R^{3/2} u_{\ast}^{1/2} K_z^{-5/6} {\rm d}z^2  + R^{3/2} u_{\ast}^{1/2} K_z^{1/6} {\rm d}\Omega_4^2 \, , 
\eeqa
where $K_z = 1 + z^2$. Considering small gauge field fluctuations ${\cal A}_\mu,{\cal A}_z$ depending only in $x^\mu$ and $z$ directions the action of the D8-\AD8 branes 
reduces to a five-dimensional $U(N_f)$ Yang Mills-Chern Simons (YM-CS) action in a curved background. The action of the model reads
\begin{align}
&S=S_{\rm YM}+S_{\rm CS} \, , \nn\\
&S_{\rm YM}=-\kappa
\int d^4 x dz\,\tr\ \left [ \frac12 K_z^{-1/3} \eta^{\mu \rho} \eta^{\nu \sigma}  {\cal F}_{\mu \nu}  {\cal F}_{\rho \sigma} 
+  M_{\ast}^2 \, K_z \eta^{\mu \nu}  {\cal F}_{\mu z}  {\cal F}_{\nu z}  \right ]\, ,\nn\\
& S_{\rm CS}= \frac{N_c}{24\pi^2}
\int_{M^4\times\bR} \tr\left(
\cA \cF^2 -\frac{i}{2}\cA^3\cF-\frac{1}{10}\cA^5
\right)\ .
\label{model}
\end{align}
where $\kappa = \lambda N_c /(216 \pi^3)$. Here, $\mu,\nu=0,1,2,3$ are four-dimensional Lorentz indices, and $z$ corresponds to the fifth dimension. The quantity
${\cA}=\cA_\alpha dx^\alpha=\cA_\mu dx^\mu+\cA_z dz
~~(\alpha=0,1,2,3,z)$ is the five-dimensional
$U(N_f)$ gauge field and $\cF=\frac{1}{2}\cF_{\alpha\beta}
dx^\alpha\wedge dx^\beta=d\cA+i\cA\wedge\cA$ is
its field strength.

\subsection{Vector meson dominance}

The gauge field ${\cal A}_\mu (x,z)$ can be expanded, in the ${\cal A}_z=0$ gauge, as
\beqa
{\cal A}_{\mu} (x, z) =  \hat {\cal V}_\mu (x)  + \hat {\cal A}_\mu (x)  \psi_0 ( z) 
+ \sum_{n=1}^{\infty} \left [ v_\mu^n (x) \psi_{2n-1} (z) +  a_\mu^n (x) \psi_{2n} (z) \right ]  \, , \label{KKexpansion} 
\eeqa
where 
\beqa
\hat {\cal V}_\mu (x)&=& \frac12  U^{-1} \left[ {\cal A}^L_{\mu} + \partial_\mu \right] U + \frac12  U \left[ {\cal A}^R_{\mu}  + \partial_\mu \right] U^{-1} \, , \cr
\hat {\cal A}_\mu (x) &=& \frac12 U^{-1} \left[ {\cal A}^L_{\mu} + \partial_\mu \right] U - \frac12  U \left[ {\cal A}^R_{\mu}  + \partial_\mu \right] U^{-1} \, , \cr
U(x) &=& e^{\frac{ i \pi(x)}{f_\pi}} \quad , \quad {\cal A}^{L(R)}_{\mu}(x) = {\cal A}_{\mu}^V (x) \pm {\cal A}_{\mu}^A (x) \, ,
\eeqa 
and the $\psi_n(z)$ modes satisfy  
\beqa
\kappa \int d z \, K_z^{-1/3} \psi_n (z) \psi_m ( z) =  \delta_{nm} \quad  ,  \quad 
- K_z^{1/3} \partial_{ z} \left[ K_z \partial_{z} \psi_n ( z) \right] = \lambda_n \,  
\psi_n (z) \, .
\eeqa

Using the Kaluza-Klein expansion (\ref{KKexpansion}) and integrating the $z$ coordinate we get a four-dimensional 
effective lagrangian of mesons and external $U(1)$ fields. 
The  vector (axial vector) mesons are represented by the fields $v_\mu^n (x)$ ($a_\mu^n (x)$) and 
correspond to the modes $\psi_{2n-1} (z)$ ($\psi_{2n} (z)$). The pion is represented by 
the field $\pi(x)$ and corresponds to the mode $\psi_0(z)$. In addition, we have external $U(1)$ vector 
(axial) fields represented by ${\cal A}_{\mu}^V$ (${\cal A}_{\mu}^A$). 

\medskip

In order to have a diagonal kinetic term, the vector mesons are redefined as 
$ \tilde v_\mu^n = v_\mu^n + (g_{v^n}/M_{v^n}^2) {\cal V}_\mu$  and the 
quadratic terms in the vector sector take the form \cite{Sakai:2005yt} :
\beqa
{\cal L}_2 &=& \frac12 \sum_n \left [ {\rm Tr} \left(\partial_\mu \tilde v_\nu^n - \partial_\nu \tilde v_\mu^n \right)^2 
+ 2  M_{v^n}^2 {\rm Tr}  \left(\tilde v_\mu^n - \frac{g_{v^n}}{M_{v^n}^2} {\cal V}_\mu \right)^2 \right ] \, , \nonumber
\eeqa
where
\beqa
 M_{v^n}^2 = \lambda_{ 2n-1} M^2_{\ast} \quad , \quad
 g_{v^n} = \kappa  \, M_{v^n}^2 \int d z \,K_z^{-1/3} \psi_{ 2n-1}(z) \, . \nonumber
\eeqa 
The mixed term $g_{v^n} \tilde v_\mu^n {\cal V}^\mu$ represents the  decay of the photon into vector mesons which is a 
holographic realization of  vector meson dominance.

\subsection{Holographic baryons}

We restrict ourselves to the case $N_f=2$. The $U(2)$ gauge field $\cA$ can be decomposed as
\begin{eqnarray}\label{eq:Adecom}
\cA=A+\wh A\,\frac{{\bf 1}_2}{2}
=A^i\frac{\tau^i}{2}+\wh A\,\frac{{\bf 1}_2}{2}
=\sum_{a=0}^3 \cA^a\,\frac{\tau^a}{2}\ ,
\end{eqnarray}
where $\tau^i$ ($i=1,2,3$) are Pauli matrices and $\tau^0={\bf 1}_2$
is a unit matrix of dimension 2.
Thus, the equations of motion are given by
\begin{eqnarray*}
&&-\kappa \left(
K_z^{-1/3} \partial_{\nu} \wh F^{\mu\nu}+ \partial_z (K_z \wh F^{\mu z})
\right)
+\frac{N_c}{128\pi^2}\epsilon^{\mu \alpha_2 \ldots \alpha_5}
\left(
F^a_{\alpha_2\alpha_3}F^a_{\alpha_4\alpha_5}
+\wh F_{\alpha_2\alpha_3}\wh F_{\alpha_4\alpha_5}
\right)
=0,\\
&& -\kappa\left(
K_z^{-1/3} \nabla_\nu F^{\mu\nu}+\nabla_z (K_z F^{\mu z})
\right)^a
+\frac{N_c}{64\pi^2}\epsilon^{\mu \alpha_2 \ldots \alpha_5}
F^a_{\alpha_2\alpha_3}\wh F_{\alpha_4\alpha_5}=0,\\
&&-\kappa
K_z \partial_{\nu} \wh F^{z\nu}
+\frac{N_c}{128\pi^2}\epsilon^{z \mu_2 \ldots \mu_5}
\left(
F^a_{\mu_2\mu_3} F^a_{\mu_4\mu_5}
+\wh F_{\mu_2\mu_3} \wh F_{\mu_4\mu_5}
\right)
=0,\\
&&-\kappa K_z
\left(\nabla_\nu F^{z\nu}\right)^a
+\frac{N_c}{64\pi^2} \epsilon^{z \mu_2 \ldots \mu_5}
F^a_{\mu_2\mu_3}\wh F_{\mu_4\mu_5}=0,
\end{eqnarray*}
where $\nabla_\alpha=\partial_\alpha+iA_\alpha$ is the covariant derivative.\\

The baryon in this model corresponds to a soliton
with a nontrivial instanton number. This is explained very well in the original paper by Sakai and Sugimoto \cite{Sakai:2004cn}, cf. also \cite{Hata:2007mb}.
The reason for this is the following: The Sakai-Sugimoto model is a holographic model of large $N_c$ QCD and naturally incorporates the Skyrme model.
It has been known for quite some time \cite{Adkins:1983ya} that the Skyrme model provides a qualitatively correct effective theory of baryons in large $N_c$ QCD.  
From symmetry considerations we can approximate the skyrmion solution by an instanton solution \cite{Atiyah:1989dq} , due to the fact that the symmetry group of the Skyrme model is $SU(2)$ 
and a skyrmion corresponds to a topological soliton solution of the pion field with homotopy $\pi_3(S^3)$.  Thus, the holographic dual of a skyrmion 
can be approximated by an instanton, i.e., a self-dual solution of the Yang-Mills equations in Euclidean space, which in our case is the four-dimensional
space parameterized by $x^M$ ($M=1,2,3,z$).

The instanton number is interpreted as the baryon number
$N_B$, where
\begin{eqnarray}
 N_B=\frac{1}{64\pi^2}\int
d^3x dz\, \epsilon_{M_1M_2M_3M_4} F^a_{M_1M_2}F^a_{M_3M_4}
\ .
\label{NB}
\end{eqnarray}
The equations of motion are complicated
nonlinear differential equations in a curved space-time, so
it is difficult to find a general analytic solution corresponding
to the baryons.
\subsubsection{Classical solution}
Since we are working in the large $\lambda$ regime, we can employ a $1/\lambda$ expansion. 
It can be easily observed that $S_{\text{CS}}$ will be subleading compared to $S_{\text{YM}}$, and therefore 
the leading contribution to the instanton mass comes from the YM action. 
As it turns out \cite{Hata:2007mb}, it is possible to focus on a small region around the center of the instanton at $z=0$ 
(because the instanton size will scale as $\lambda^{-1/2}$), where the warp factor $K_z$ is approximately 1. The corresponding field equations will be solved by a BPST instanton with infinitesimal size $\rho \rightarrow 0$. 
As explained in \cite{Hata:2007mb}, including the contributions to the field equations from the CS term will
induce a non-vanishing $U(1)$ electric field $\wh{A}_0$ and will stabilize the size of the instanton (determined by the minimum of the effective potential for $\rho$) at a finite value.
The classical solution near $z=0$ corresponds  to a static baryon configuration and is given by
 \begin{align}
A_M^{\rm cl}=&-if(\xi)g\del_M g^{-1} \ ,~~
\wh A_0^{\rm cl}=
\frac{N_c}{8\pi^2 \kappa}
\frac{1}{\xi^2}
\left[1-\frac{\rho^4}{(\rho^2+\xi^2)^2}
\right]\ ,~~~
A_0=\wh A_M=0 .
\label{eq:clBS}
\end{align}
with the definitions
\begin{equation}
f(\xi)=\frac{\xi^2}{\xi^2+\rho^2}\ ,~~~
g(x)=\frac{(z-Z)-i(\vec{x}-\vec{X})\cdot\vec\tau}{\xi} \ ,~~
\xi= \sqrt{(z-Z)^2+|\vec{x}-\vec{X}|^2}  ,
\end{equation}
where $X^M=(X^1,X^2,X^3,Z)=(\vec X,Z)$ determines the position in the spatial $\bR^4$ direction. The effective potential for $\rho$ and $Z$ can be calculated by taking into account the nontrivial $z$-dependence of the background (through $K_z$) at order $\lambda^{-1}$ and reads
\begin{equation}
 V_{\text{eff}}(\rho,Z)= M_0
\left(
1+\frac{\rho^2}{6}
+\frac{N_c^2}{5 M_0^2}\frac{1}{\rho^2}
+\frac{Z^2}{3}
\right) ,
\end{equation}
where $M_0=8\pi^2\kappa M_{\ast}$ is the minimal (groundstate) mass of the baryons.
The effective potential is minimized at
\begin{equation}
\rho_{\rm cl}^2
=\frac{N_c}{M_0}\sqrt{\frac{6}{5}}\ ,
~~
Z_{\rm cl}=0.
\end{equation}
\subsubsection{Quantization}
The quantization of the solitons is facilitated by employing the moduli space approximation method to study a quantum mechanical problem on the instanton moduli space. 
For a more detailled discussion, the interested reader is referred to refs. \cite{Hata:2007mb,Hashimoto:2008zw}. 
Here we merely present the results for the wavefunctions and energies of the lowest excited baryon states in the slowly moving (pseudo-) moduli approximation. These (pseudo-) moduli are:
\begin{eqnarray}
 X^i(t)\  ,~Z(t)\  ,~ \rho(t)\  ,~a^I(t)\ , \label{coleccord}
\end{eqnarray}
where $X^i$ and $Z$ represent the center-of-mass position of the
soliton, while $\rho$ is the size of the instanton and $a^I$ ($I=1,2,3,4$) determines the orientation of the instanton in the $SU(2)$ group space,
with the condition $(a^I)^2=1$. 
The $SU(2)$ gauge field takes the form 
\beqa
A_M = V A_M^{\rm cl} V^{-1} - i V \partial_M V^{-1}  \, , \label{timedepsol}
\eeqa
where $A_M^{\rm cl}$ is given by eq. (\ref{eq:clBS}) and  $V$ satisfies the Gauss law constraint
\beqa
- i V^{-1} \dot V = - \dot X^M A^{\rm cl}_M + \chi^a f(\xi) g \frac{\tau^a}{2} g^{-1} \, , 
\eeqa
with 
\beqa
\chi^a = - i \tr ( \tau^a {\bf a}^{-1} \dot {\bf a} ) \quad , \quad 
{\bf a} = a_4 + i a_a \tau^a \, .
\eeqa
Inserting (\ref{timedepsol}) into the effective action we get the Lagrangian of collective motion of the instanton 
\beqa
L = \frac{M_0}{2} (  \dot{\vec{X}}^2 + \dot Z^2 ) + M_0 \sum_{I=1}^4 \dot{(\rho a_I)} \dot{(\rho a_I)}  -  V_{\text{eff}}(\rho,Z) \, .
\eeqa
Quantizing this system we find the baryon wavefunctions as eigenstates of the Hamiltonian. The relevant quantum numbers are $B=(l,I_3,n_{\rho}, n_z)$ and its spin $s$.
For example, baryon wavefunctions with $B_n=(1, + 1/2,0,n)$ can be written as
\begin{eqnarray}
 \ket{\, B_n \uparrow\,}\propto
R(\rho)\psi_{B_n}(Z)(a_1+ia_2)\, , \label{Barstate}
\end{eqnarray}
where  
\beqa
 R(\rho) &=& \rho^{-1+2\sqrt{1+N_c^2/5}}
e^{-\frac{M_0}{\sqrt{6}}\rho^2}\, ,  \cr
\Psi_{B_n} (Z) &=& \left(\frac{(2 M_0)^{1/4}}{6^{1/8}\,\pi^{1/4}\,2^{n/2}\,\sqrt{n!}}\right)H_n(\sqrt{2 M_0} 6^{-1/4} Z) \text{e}^{-\frac{M_0}{\sqrt{6}}Z^2}. 
\eeqa
Note that eq. (\ref{Barstate}) gives a representation of a baryonic state in terms of the collective coordinates (\ref{coleccord}) of the moduli space of the instanton.
The baryon masses can be easily gleaned  from the relevant Hamiltonians and the resulting mass formula reads,
 \begin{equation}\label{eq:masses}
 M = M_0 + \sqrt{\frac{(\ell+1)^2}{6}+\frac{2}{15}N_c^2}+\frac{2(n_{\rho}+n_z)+2}{\sqrt{6}}=: {\wt M}_0 + \frac{2 n_z}{\sqrt{6}}.
 \end{equation}
One observation is in order: Since the photon couples to the baryons via vector mesons as a direct consequence of vector meson dominance and since the
 vector meson wavefunctions only depend on the coordinate $z$, it is clear that the intial and final state baryons must have the same $\rho$ quantum number due 
to orthonormality, while they may differ in the $z$ quantum number, due to the additional contribution from the vector mesons to the relevant coupling constants etc.

\subsubsection{Extension of the soliton solution to large $z$}

The classical solution (\ref{eq:clBS}) is valid only near $z=0$. This solution can be extended to large $z$ as long as we require $\rho \ll \xi$ which is the condition
of small size for the skyrmion. Under this condition the equations of motion linearize and the solutions can be found by defining Green's functions corresponding to the 
curved space generated by $K_z$ :
\beqa
G( \vec{x}, z , \vec{X}, Z) &=& \kappa \sum_{n=1}^\infty \psi_n(z) \psi_n(Z) Y_n (| \vec{x} - \vec{X}|) \cr
H( \vec{x}, z , \vec{X}, Z) &=& \kappa \sum_{n=0}^\infty \phi_n(z) \phi_n(Z) Y_n (| \vec{x} - \vec{X}|) \, ,
\eeqa
where $\psi_n(z)$ is the complete set of vector meson eigenfunctions, and $\phi_n(z)$ is another set defined by  
\beqa
\phi_0(z) = \frac{1}{\sqrt{\kappa \pi} K_z}  \quad , \quad \phi_n(z) = \frac{1}{\sqrt{\lambda_n}} \partial_z \psi_n (z) \quad (n=1,2,\dots) \, ,
\eeqa
and $Y_n(r)$ is the Yukawa potential
\beqa
Y_n(r) = - \frac{1}{4 \pi} \frac{e^{- \sqrt{\lambda_n} r}} {r} \, .
\eeqa
The gauge field solutions found in \cite{Hashimoto:2008zw} for the case  $\rho \ll \xi$ can be written as 
\beqa
\hat A_0 &=& - \frac{N_c}{2 \kappa} G( \vec{x}, z , \vec{X}, Z) \, ,\cr
\hat A_i &=& \frac{N_c}{2 \kappa} \left \{ \dot X^i + 
\frac{\rho^2}{2} \left[ \frac{\chi^a}{2} \left( \epsilon^{iaj} \frac{\partial}{\partial X^j} - \delta^{ia} \frac{\partial}{\partial Z} \right)
+ \frac{ \dot \rho}{\rho}  \frac{\partial}{\partial X^i} \right ] \right \} G( \vec{x}, z , \vec{X}, Z)  \, , \cr
\hat A_z &=& \frac{N_c}{2 \kappa}  \left [ \dot Z + \frac{\rho^2}{2} \left ( \frac{\chi^a}{2} \frac{\partial}{\partial X^a} + 
\frac{ \dot \rho}{\rho}  \frac{\partial}{\partial Z} \right )  \right ] H( \vec{x}, z , \vec{X}, Z)  \, , \cr
A_0^\Lambda &=& 2 \pi^2 \rho^2 \Big \{ 2 i  \, {\bf a} \, \dot {\bf a}^{-1} 
+ 2 \pi^2 \rho^2 {\bf a} \tau^a {\bf a}^{-1} 
\left [ \dot X^i \left( \epsilon^{iaj} \frac{\partial}{\partial X^j} - \delta^{ia} \frac{\partial}{\partial Z} \right) + \dot Z \frac{\partial}{\partial X^a} \right ]
\Big \}  G( \vec{x}, z , \vec{X}, Z) \, , \cr
A_i^\Lambda&=& - 2 \pi^2 \rho^2 {\bf a} \tau^a {\bf a}^{-1} \left( \epsilon^{iaj} \frac{\partial}{\partial X^j} - \delta^{ia} \frac{\partial}{\partial Z} \right) 
 G( \vec{x}, z , \vec{X}, Z) \, , \cr 
A_z^\Lambda &=& - 2 \pi^2 \rho^2 {\bf a} \tau^a {\bf a}^{-1} \frac{\partial}{\partial X^a} H( \vec{x}, z , \vec{X}, Z) \label{Az} \, , 
\eeqa
where
\beqa
A_\alpha^\Lambda = \Lambda A_\alpha \Lambda^{-1} - i \Lambda \partial_\alpha \Lambda \quad , \quad 
\Lambda = {\bf a} g^{-1} V^{-1} \, .
\eeqa

\section{Generalized baryon form factors in the Sakai-Sugimoto model }\label{sec:GFF}

Now we are going to extract the generalized Dirac and Pauli form factors by comparing the  matrix element of the vectorial current in (\ref{matrixelement}) with the 
one that can be obtained from the Sakai-Sugimoto model. The latter, denoted here by $J_{V (SS)}^{\mu,a}$, can be gleaned from holography as \cite{Hashimoto:2008zw} 
\beqa
J_{V (SS)}^{\mu,a} = - \kappa \Big \{ \lim_{z \to \infty} \left [ K_z {\cal F}^{\text{cl}}_{\mu z} \right ]
 + \lim_{z \to - \infty} \left [ K_z {\cal F}^{\text{cl}}_{\mu z} \right ] \Big \} \,, \label{holcurrent}
\eeqa
where ${\cal F}^{\text{cl}}_{\mu z}$ is the field strength associated with the classical field (\ref{Az}).
When comparing the vectorial current of (\ref{matrixelement}) with the one in the Sakai-Sugimoto model 
we will use the following prescription
\beqa
 \eta_\mu \langle p_{  X} , B_{  X} , s_{  X}  \vert J_V^{\mu,a} (0) \vert p , B, s \rangle 
= \eta_\mu \langle p_{  X} , B_{  X} , s_{  X}  \vert J_{V (SS)}^{\mu,a} (0) \vert p , B, s \rangle \label{holprescription}
\eeqa
where $\eta_\mu = (\eta_0 , \vec{\eta})$ is the polarization of the photon and we choose to work with transverse photons satisfying the relation $\eta_\mu q^\mu=0$
in order to avoid the discussion of current anomalies. 

\subsection{Electromagnetic currents in the Sakai-Sugimoto model}

Let us now represent the electromagnetic currents as operators  in the moduli space of the collective coordinates of the instanton. 
Using (\ref{Az}) and (\ref{holcurrent})  one gets \cite{Hashimoto:2008zw} : 
\beqa
J^{0,0}_{V (SS)} (x)&=& \frac{N_c}{2} G_V \, , \cr
J^{i,0}_{V (SS)} (x)&=& - \frac{N_c}{2} \Big \{ \dot Z \partial^i H_V - \dot X^i G_V
- \frac{S_a}{16 \pi^2 \kappa} \left[ (\partial^i \partial^a - \delta^{ia} \partial^2 ) H_V
+ \epsilon^{ija} \partial_j G_V \right ] \Big \} \, ,\cr
J^{0,c}_{V (SS)} (x)&=& 2 \pi^2 \kappa \Big \{ \rho^2 \tr [ \tau^c \partial_0 (  {\bf a} \tau^a {\bf a}^{-1} ) ] \partial_a H_V
+ \frac{I^c}{2 \pi^2 \kappa} G_V  \cr
&-& \rho^2 \tr [ \tau^c  {\bf a} \tau^a {\bf a}^{-1} ] \dot X^i \left[ (\partial_a \partial_i - \delta_{ia} \partial^2 )H_V
+ \epsilon^{ija} \partial_j G_V \right ]  \Big \} \, , \cr
J^{i,c}_{V (SS)} (x)&=& - 2 \pi^2 \kappa \rho^2 \tr [ \tau^c  {\bf a} \tau_a {\bf a}^{-1} ]
\left [ (\partial^i \partial^a - \delta^{ia} \partial^2 ) H_V + \epsilon^{ija} \partial_j G_V \right ] \,,
\eeqa
where
\beqa
G_V &=& - \sum_n g_{v^n} \psi_{2n-1}(Z) Y_{2n-1}(|\vec{x} - \vec{X}|) \, , \cr
H_V &=& - \sum_n \frac{g_{v^n}}{\lambda_{2n-1}} \partial_Z \psi_{2n-1}(Z) Y_{2n-1}(|\vec{x} - \vec{X}|) \,, \cr
\dot Z &=& - \frac{i}{M_0} \partial_Z =  \frac{P_Z}{M_0} \quad , \quad
\dot X^i = - \frac{i}{M_0} \frac{\partial}{\partial X^i} = \frac{P^i}{M_0} \, ,
\eeqa
and
\beqa
S^a = 4 \pi^2 \kappa \rho^2 \chi_a = - i 4 \pi^2 \kappa \rho^2 \tr ( \tau^a {\bf a}^{-1} \dot {\bf a}) \quad , \quad
I^a =  - i 4 \pi^2 \kappa \rho^2 \tr ( \tau^a {\bf a} \dot {\bf a}^{-1} ) \, ,
\eeqa
are the spin and isospin operators. Note that
\beqa
\dot Z (\partial^i H_V) - \dot X^i G_V &=&
\frac{1}{M_0} \Big [ (\partial^i H_V) P_Z - G_V P^i \Big ] \, .
\eeqa
Here we used the relation $\partial_Z H_V = - G_V$.\\
Defining the Fourier transform as
\beqa
\tilde J_{V (SS)}^{\mu, a}(\vec{k}) = \int d^3 \vec{x} e^{- i \vec{k} \cdot x} J_{V (SS)}^{\mu, a} (x) \, ,
\eeqa
and using the identity
\beqa
\int d^3 \vec{x} e^{-i \vec{k} \cdot x} Y_{2n-1}(|\vec{x} - \vec{X}|)
= - \frac{e^{- i \vec{k} \cdot \vec{X}}}{\vec{k}^2 + \lambda_{2n-1} } \, ,
\eeqa
we find
\beqa
\tilde J^{0,0}_{V (SS)}(\vec{k}) = \frac{N_c}{2} e^{-i \vec{k} \cdot \vec{X}} \sum_n \frac{g_{v^n} \psi_{2n-1}(Z)}{\vec{k}^2 + \lambda_{2n-1}} \, ,
\eeqa
\beqa
\tilde J^{i,0}_{V (SS)}(\vec{k}) &=& \frac{N_c}{2} e^{-i \vec{k} \cdot \vec{X}} \Big \{
\sum_n \frac{g_{v^n} \psi_{2n-1}(Z)}{\vec{k}^2 + \lambda_{2n-1}}
\left[ \frac{P^i}{M_0} + \frac{i}{16 \pi^2 \kappa} \epsilon^{ija} k_j S_a \right ]  \cr
&-& \sum_n \frac{g_{v^n} \partial_Z \psi_{2n-1}(Z)}{\lambda_{2n-1} (\vec{k}^2 + \lambda_{2n-1})}
\left [ \frac{k^i}{M_0} \partial_Z + \frac{1}{16 \pi^2 \kappa } (k^i k^a - \vec{k}^2 \delta^{ia} ) S_a \right ] \Big \} \, ,
\eeqa
\beqa
\tilde J^{0,c}_{V (SS)}(\vec{k}) &=& 2 \pi^2 \kappa \, e^{-i \vec{k} \cdot \vec{X}} \Big \{
\sum_n \frac{g_{v^n} \psi_{2n-1}(Z)}{\vec{k}^2 + \lambda_{2n-1}}
\left [ \frac{I^c}{2 \pi^2 \kappa}
- \frac{i}{M_0} \epsilon^{ija} P_i k_j \rho^2 \tr (\tau^c  {\bf a} \tau_a {\bf a}^{-1} ) \right ]  \cr
&+& \sum_n \frac{g_{v^n} \partial_Z \psi_{2n-1}(Z)}{\lambda_{2n-1} (\vec{k}^2 + \lambda_{2n-1})}
 \left [  i k_i \rho^2 \tr [ \tau^c \partial_0 (  {\bf a} \tau^i {\bf a}^{-1}  )]
+ \frac{1}{M_0} (\vec{P} \cdot \vec{k} k_i - \vec{k}^2 P_i )
\rho^2 \tr [ \tau^c  {\bf a} \tau^i {\bf a}^{-1}  ] \right ] \Big \} \, , \cr
&&
\eeqa
\beqa
\tilde J^{i,c}_{V (SS)}(\vec{k}) &=& 2 \pi^2 \kappa \, e^{-i \vec{k} \cdot \vec{X}} \Big [
- i \sum_n \frac{g_{v^n} \psi_{2n-1}(Z)}{\vec{k}^2 + \lambda_{2n-1}}
\epsilon^{ija} k_j
\cr
&+& \sum_n \frac{g_{v^n} \partial_Z \psi_{2n-1}(Z)}{\lambda_{2n-1} (\vec{k}^2 + \lambda_{2n-1})}
(k^i k^a - \vec{k}^2 \delta^{ia} ) \Big ] \rho^2 \tr ( \tau^c  {\bf a} \tau_a {\bf a}^{-1} ) \,.
\eeqa
Note that one term arising from $\dot Z$ cancels with another from $\dot X^i$ and we have used the relation $\partial_Z^2 \psi_n (Z) \approx  -\lambda_n \psi_n (Z)$.
Now we calculate the expectation values of the Sakai-Sugimoto currents :
\beqa
\langle p_{ X} , B_{ X} , s_{ X} \vert J^{\mu,a}_{V (SS)} (0) \vert  p , B,  s \rangle = \int  \frac{d^3 \vec{k}}{(2 \pi)^3} \langle p_{ X} , B_{ X} , s_{ X} \vert \tilde J^{\mu,a}_{V (SS)} (\vec{k})
\vert  p , B , s \rangle \, .
\eeqa
We define the baryon states as 
\beqa
\vert \vec{p}, B,   s  , I_3 \rangle &=&  \frac{1}{(2 \pi)^{3/2}} e^{ i \vec{p} \cdot \vec{X}} \vert n_B \rangle \vert n_\rho \rangle  \vert  s , I_3 \rangle_R \, , \cr
\vert \vec{p}_{ X}, B_{ X} , s_{ X} , I_3^{ X}  \rangle  &=&  \frac{1}{(2 \pi)^{3/2}} e^{ i \vec{p}_{ X} \cdot \vec{X}} \vert n_{B_{ X}} \rangle \vert n_\rho \rangle 
\vert  s_{ X} , I_3^{ X}  \rangle_R \, .
\eeqa
Here we make use of the results and definitions of a recent publication \cite{BoschiFilho:2011hn}, in which a relativistic generalization of baryon states and wavefunctions was discussed in detail. 
In particular, the spin and isospin part was defined as 
\beqa
\vert  s , I_3 \rangle_{ R} &=&  \frac{1}{\sqrt{2E}} \left( \begin{array}{c} f \, \vert  s , I_3 \rangle \\ \frac{ s |\vec{p}| }{f} \, \vert  s , I_3 \rangle \end{array} \right) \, , \cr  
\langle  s_{ X} , I_3^{ X}  \vert_{ R} &=&  \frac{1}{\sqrt{2 E_{ X}}} \left(  f_{ X} \, \langle  s_{ X} , I^{ X}_3 \vert \quad  - \frac{ s_{ X}|\vec{p}_{ X}| }{f_{ X}} 
\, \langle  s_{ X} , I^{ X}_3 \vert  \right) \, , 
\eeqa
where $\vert  s , I_3 \rangle$ and $\langle  s_{ X} , I^{ X}_3 \vert$ are the non-relativistic initial and final states associated with the spin and isospin operators.
Evaluating the currents in these states we get 
\beqa
\langle J^{0,0}_{V (SS)}(0) \rangle =  \frac{1}{(2 \pi)^3} \frac{N_c}{2}  \langle s_{ X} , I_3^{ X}  \vert s , I_3 \rangle_{ R}   F^1_{B B_{ X}}(\vec{q}^2) \, ,
\eeqa
\beqa
\langle J^{i,0}_{V (SS)}(0) \rangle &=& \frac{1}{(2 \pi)^3} \frac{N_c}{2}  \langle s_{ X} , I_3^{ X} \vert_{ R} \Big \{
F^1_{B B_{ X}}(\vec{q}^2) \left[ \frac{p^i}{M_0}   - \frac{i}{16 \pi^2 \kappa} \epsilon^{ija} q_j S_a \right ]  \cr
&+& \frac{q^i}{M_0} F^3_{B B_{ X}}(\vec{q}^2)   - \frac{1}{16 \pi^2 \kappa } F^2_{B B_{ X}} (\vec{q}^2)(q^i q^a - \vec{q}^2 \delta^{ia} ) S_a  \Big \} \vert s , I_3 \rangle_{ R} \, ,
\eeqa
\beqa
\langle J^{0,c}_{V (SS)}(0) \rangle  &=& 2 \pi^2 \kappa \,  \frac{1}{(2 \pi)^3} \langle  n_\rho \vert \langle  s_{ X} , I_3^{ X}  \vert_{ R} \Big \{
F^1_{B B_{ X}}(\vec{q}^2) \left [ \frac{ I^c    }{2 \pi^2 \kappa}
+ \frac{i}{M_0} \epsilon^{ija} p_i q_j  \rho^2 \tr (\tau^c  {\bf a} \tau_a {\bf a}^{-1} )   \right ]  \cr
&+& F^2_{B B_{ X}}(\vec{q}^2) \Big [ - i q_i  \rho^2 \tr [ \tau^c \partial_0 (  {\bf a} \tau^i {\bf a}^{-1}  )  ] \cr
&+& \frac{1}{M_0} (\vec{P} \cdot \vec{q} q_i - \vec{q}^2 P_i )
 \rho^2 \tr [ \tau^c  {\bf a} \tau^i {\bf a}^{-1}  ]  \Big ]  \Big \} \vert n_\rho \rangle \vert  s , I_3  \rangle_{ R} \, , 
\eeqa
\beqa
\langle J^{i,c}_{V (SS)}(0) \rangle &=&  2 \pi^2 \kappa \, \frac{1}{(2 \pi)^3} \Big [
 i F^1_{B B_{ X}}(\vec{q}^2) \epsilon^{ija} q_j  + F^2_{B B_{ X}}(\vec{q}^2) (q^i q^a - \vec{q}^2 \delta^{ia} ) \Big ] \cr 
&\times& \langle n_\rho \vert \rho^2 \vert n_\rho \rangle \langle  s_{ X} , I_3^{ X}  \vert_{ R}  \tr ( \tau^c  {\bf a} \tau_a {\bf a}^{-1} ) \vert  s , I_3  \rangle_{ R} \,.
\eeqa
where
\beqa
F^1_{B B_{ X}}(\vec{q}^2) &=& \sum_n \frac{g_{v^n} \langle n_{B_{ X}} \vert \psi_{2n-1}(Z) \vert n_B \rangle}{\vec{q}^2 + \lambda_{2n-1}}  \cr
F^2_{B B_{ X}}(\vec{q}^2) &=& \sum_n \frac{g_{v^n} \langle n_{B_{ X}} \vert \partial_Z \psi_{2n-1}(Z)  \vert n_B \rangle }{\lambda_{2n-1} (\vec{q}^2 + \lambda_{2n-1})}  \cr
F^3_{B B_{ X}}(\vec{q}^2) &=&\sum_n \frac{g_{v^n} \langle n_{B_{ X}} \vert \partial_Z \psi_{2n-1}(Z) \partial_Z  \vert n_B \rangle }{\lambda_{2n-1} (\vec{q}^2 + \lambda_{2n-1})} \,, \label{F1F2F3}
\eeqa
the momentum $\vec{q}$ is the photon momentum defined by $\vec{q} = \vec{p}_{ X} - \vec{p}$ and we have used 
\beqa
\langle \vec{p}_{ X} \vert e^{- i \vec{k} \cdot \vec{X}} \vert \vec{p} \rangle
=  \delta^3 ( \vec{k} - \vec{p} + \vec{p}_{ X}) \,. \label{ident}
\eeqa
The following relations are very useful and are presented here for completeness,
\beqa
 \langle s_{ X} , I_3^{ X}  \vert s , I_3 \rangle_{ R} &=& 
\frac{1}{2\sqrt{E_XE}} (ff_X-\tfrac{ss_{ X}|p||\vec{p}_{  X}|}{ff_X})\,  \delta_{I_3^{ X} I} \chi^\dagger_{s_{ X}}(\vec{p_{ X}})\, \chi_s (\vec{p})\, ,\cr \nonumber
  \langle s_{ X} , I_3^{ X} \vert_{ R} \tr  ( \tau^c  {\bf a} \tau_a^{-1} {\bf a} ) \vert s , I_3 \rangle_{ R} &=& 
-\frac{1}{3\sqrt{E_XE}} (ff_X-\tfrac{ss_{ X}|p||\vec{p}_{  X}|}{ff_X})\, \tau^c_{I_3^X I_3} \, \chi^\dagger_{s_{ X}}(\vec{p_{ X}})\,\sigma^a \chi_s (\vec{p})\, , \cr
\langle s_{ X} , I_3^{ X}  \vert_{ R} \, I^c \, \vert  s , I_3 \rangle_{ R} &=& 
\frac{1}{4 \sqrt{E_X E}} (ff_X-\tfrac{ss_{ X}|p||\vec{p}_{  X}|}{ff_X})\,  (\tau^c)_{I_3^{ X} I} \chi^\dagger_{s_{ X}}(\vec{p_{ X}})\, \chi_s (\vec{p})\, ,\cr
\langle s_{ X} , I_3^{ X}  \vert_{ R} \, S_a \, \vert  s , I_3 \rangle_{ R} &=& 
\frac{1}{4 \sqrt{E_X E}} (ff_X-\tfrac{ss_{ X}|p||\vec{p}_{  X}|}{ff_X})\,  \delta_{I_3^{ X} I} \chi^\dagger_{s_{ X}}(\vec{p_{ X}})\, \sigma_a \chi_s (\vec{p})\, ,\cr
\eeqa

{\bf Positive parity resonances in the Breit frame.}
In the Breit frame we get for positive parity resonances
\beqa
\langle J^{0,0}_{V (SS)}(0) \rangle &=&  \frac{N_c}{2 (2 \pi)^3} \xi  \delta_{I_3^{ X} I} \chi^\dagger_{s_{ X}}(\vec{p_{ X}}) \chi_s (\vec{p})  F^1_{B B_{ X}}(\vec{q}^2) \, , \cr
\langle J^{i,0}_{V (SS)}(0) \rangle &=& \frac{N_c}{2 (2 \pi)^3 M_0}   \delta_{I_3^{ X} I} \chi^\dagger_{s_{ X}}(\vec{p_{ X}}) \Big \{
 q^i \left [ F^3_{B B_{ X}}(\vec{q}^2) - \frac{1}{2x} F^1_{B B_{ X}}(\vec{q}^2) \right ] \xi  \cr 
&-& \frac{i}{4} \alpha \epsilon^{ija} q_j \sigma_a F^1_{B B_{ X}}(\vec{q}^2)  \Big \} \chi_s(\vec{p}) \, , \cr
\langle J^{0,c}_{V (SS)}(0) \rangle  &=& \frac{\xi}{2 (2 \pi)^3} ( \tau^c )_{I_3^{ X} I}  \chi^\dagger_{s_{ X}}(\vec{p_{ X}}) \chi_s (\vec{p})  F^1_{B B_{ X}}(\vec{q}^2)  \, , \cr
\langle J^{i,c}_{V (SS)}(0) \rangle &=&  - i \frac{\alpha}{2 (2 \pi)^3} \left ( \frac{M_0}{3} \right ) (\tau^c)_{I_3^{ X} I} \langle n_\rho \vert \rho^2 \vert n_\rho \rangle \,  
   \epsilon^{ija} q_j  \chi^\dagger_{s_{ X}}(\vec{p_{ X}}) \sigma_a  \chi_s (\vec{p})  F^1_{B B_{ X}}(\vec{q}^2) \, , \, \label{SSCurrentBreit}
\eeqa
where 
\beqa
\xi &=& \left (\frac{1}{2E} \right ) \left ( \frac{\sqrt{E + m_B}}{\sqrt{E + m_{B_{ X}}}} \right ) \left [ E + m_{B_{ X}} + (E - m_B) (2x-1) \right ] \, , \cr
 \left ( \frac{M_0}{3} \right ) \langle n_\rho \vert \rho^2 \vert n_\rho \rangle &=& 
\frac{1}{\sqrt{6} M_{\ast} } \left [ 1 + 2 \sqrt{ 1 + \frac{N_c^2}{5} }  \right ] \equiv \frac{g_{I=1}}{4 m_B}  \,. \label{xidef}
\eeqa
and  $\alpha$ is given in (\ref{alphabeta}). We have also used the relation
\beqa
(ff_X-\tfrac{ss_{ X}|p||\vec{p}_{  X}|}{ff_X}) = \frac{f}{f_X} \left[E+m_{B_X} -s s_{ X} (E-m_B)\vert 2x-1 \vert \right] \, ,
\eeqa
and the identity (\ref{helident}) which are valid only in the Breit frame. 

\subsection{The generalized Dirac and Pauli form factors}

Using the holographic prescription (\ref{holprescription}) we can compare, in the case of positive parity baryons, 
the current matrix element of (\ref{CurrentBreit0}),(\ref{CurrentBreiti}) with the Sakai-Sugimoto current matrix element in 
(\ref{SSCurrentBreit}). As a consequence we get our main result: the Dirac and Pauli form factors 
\beqa
F^{D,0}_{B B_{  X}}(q^2) &=& \left [ \frac{\xi \alpha + \beta \alpha \frac{q^2 }{4 M_0}  }{\alpha^2 + \beta^2 q^2} \right ] N_c F^1_{B B_{ X}}(q^2) \, ,
\eeqa

\beqa
F^{P,0}_{B B_{  X}}(q^2) &=& - \frac{1}{\kappa_B} \left [ \frac{\beta \xi - \frac{\alpha^2}{4 M_0}  }{\alpha^2 + \beta^2 q^2 } \right ]  N_c F^1_{B B_{ X}}(q^2) \, ,
\eeqa

\beqa
F^{D,3}_{B B_{  X}}(q^2) &=& \left [ \frac{\xi \alpha + \beta \alpha q^2  \left ( \frac{M_0}{3} \right ) \langle \rho^2 \rangle  }{\alpha^2 + \beta^2 q^2} \right ]   F^1_{B B_{ X}}(q^2) \, ,
\eeqa

\beqa
F^{P,3}_{B B_{  X}}(q^2) &=& - \frac{1}{\kappa_B} \left [ \frac{\beta \xi - \alpha^2 \left ( \frac{M_0}{3} \right ) \langle \rho^2 \rangle }{\alpha^2 + \beta^2 q^2 } \right ]   F^1_{B B_{ X}}(q^2) \, ,
\eeqa

\noindent where $\alpha$ and $\beta$ are given in (\ref{alphabeta}) and $\xi$ is given in (\ref{xidef}). It is important to remark that the generalized Dirac and Pauli form factors depend on the collective coordinates 
of the instanton through eq. (\ref{F1F2F3}) and the expected value $\langle \rho^2 \rangle$.

The electromagnetic Dirac and Pauli form factors read 
\beqa
F^{D}_{B B_{  X}}(q^2) &=& \frac12 \left [ \frac{1}{N_c} F^{D,0}_{B B_{  X}}(q^2) + F^{D,3}_{B B_{  X}}(q^2) \right ] \, , \cr 
F^{P}_{B B_{  X}}(q^2) &=& \frac12 \left [ \frac{1}{N_c} F^{P,0}_{B B_{  X}}(q^2) + F^{P,3}_{B B_{  X}}(q^2) \right ] \, .
\eeqa

\subsubsection{The non-relativistic (large $\lambda$) limit}

First of all note that 
\beqa
m_{B_{ X}} = m_B + \frac{2}{\sqrt{6}} n_{B_X} M_{\ast} \, ,
\eeqa
so that 
\beqa
 x = \frac{q^2}{q^2 + \left ( 2 m_B + \frac{2}{\sqrt{6}} n_{B_X} M_{\ast}   \right ) \left ( \frac{2}{\sqrt{6}} n_{B_X} M_{\ast} \right )} \,.
\eeqa
{\bf Elastic case.} In the elastic case we have  $n_{B_X}=0$, $m_{B_{ X}}=m_B$ and $x=1$ so in the large $\lambda$ limit we find
\beqa
- \frac{1}{\kappa_B} \left ( \beta - \frac{\alpha}{4 M_0} \right ) &=& \frac{m_B}{2 M_0} - 1 + {\cal O} \left ( \frac{1}{ \lambda N_c} \right )
 = \frac{g_{I=0}}{2} - 1 + {\cal O} \left ( \frac{1}{ \lambda N_c} \right ) \, ,\cr 
- \frac{1}{\kappa_B} \left [ \beta - \alpha \left ( \frac{M_0}{3} \right )  \langle \rho^2 \rangle \right ] 
&=& \frac{g_{I=1}}{2} \left [ 1 + {\cal O} \left ( \frac{1}{ \lambda N_c} \right ) \right ] \,, 
\eeqa
so that 

\beqa
F^{D,0}_{B B_{  X}}(q^2) &=& \left [ 1 + {\cal O} \left ( \frac{1}{ \lambda^2 N_c^2} \right )  \right ]  N_c F^1_{B B_{ X}}(q^2) \, ,
\eeqa

\beqa
F^{P,0}_{B B_{  X}}(q^2) &=&  \left [  \frac{g_{I=0}}{2} - 1 + {\cal O} \left ( \frac{1}{ \lambda N_c} \right ) \right ]  N_c F^1_{B B_{ X}}(q^2) \, ,
\eeqa

\beqa
F^{D,3}_{B B_{  X}}(q^2) &=& \left [ 1 + {\cal O} \left ( \frac{1}{\lambda} \right ) \right ]   F^1_{B B_{ X}}(q^2) \, ,
\eeqa

\beqa
F^{P,3}_{B B_{  X}}(q^2) &=& \frac{g_{I=1}}{2} \left [ 1 + {\cal O} \left ( \frac{1}{ \lambda N_c} \right ) \right  ]   F^1_{B B_{ X}}(q^2) \, ,
\eeqa
{\bf Non-elastic case.} In the non-elastic case, we have $n_{B_X}={2,4,6,\dots}$, so that 
in the large $\lambda$ limit we have 
\beqa
\frac{m_B}{E} &=&\left( 1+ \frac{q^2}{4 x^2 m_B^2} \right)^{-1/2}=\left(1 + \frac23 \frac{ n_{B_X}^2 M_{\ast}^2 }{q^2}\right)^{-1/2} \sim {\cal O}(1) \, , \cr
\xi \alpha + \beta \alpha \frac{q^2}{4 M_0} &=& \frac{m_B}{E} + {\cal O} \left ( \frac{1}{ \lambda N_c} \right ) \, , \cr 
- \frac{1}{\kappa_B} \left ( \beta \xi - \frac{\alpha^2}{4 M_0} \right ) &=& 
\frac{m_B}{2 M_0} - \frac{m_B}{E}
 + {\cal O} \left ( \frac{1}{ \lambda N_c} \right ) \cr 
 &=& \frac{g_{I=0}}{2} -\frac{m_B}{E}
 + {\cal O} \left ( \frac{1}{ \lambda N_c} \right ) \cr 
\xi \alpha + \beta \alpha q^2  \left ( \frac{M_0}{3} \right )  \langle \rho^2 \rangle  &=& \frac{m_B}{E} + {\cal O} \left ( \frac{1}{\lambda} \right ) \, , \cr
- \frac{1}{\kappa_B} \left [ \beta \xi - \alpha^2 \left ( \frac{M_0}{3} \right )  \langle \rho^2 \rangle \right ] 
&=& \frac{g_{I=1}}{2} \left [ 1 + {\cal O} \left ( \frac{1}{ \lambda N_c} \right )\right] \,.
\eeqa
Thus, in the non-elastic case, we obtain in the large $\lambda$ limit 
\beqa
F^{D,0}_{B B_{  X}}(q^2) &=& \left [ \frac{m_B}{E} + {\cal O} \left ( \frac{1}{ \lambda N_c} \right )  \right ]  N_c F^1_{B B_{ X}}(q^2) \, ,
\eeqa

\beqa
F^{P,0}_{B B_{  X}}(q^2) &=&  \left [  \frac{g_{I=0}}{2} -\frac{m_B}{E} + {\cal O} \left ( \frac{1}{ \lambda N_c} \right ) \right ]  N_c F^1_{B B_{ X}}(q^2) \, ,
\eeqa

\beqa
F^{D,3}_{B B_{  X}}(q^2) &=& \left [ \frac{m_B}{E} + {\cal O} \left ( \frac{1}{\lambda} \right ) \right ]   F^1_{B B_{ X}}(q^2) \, ,
\eeqa

\beqa
F^{P,3}_{B B_{  X}}(q^2) &=& \frac{g_{I=1}}{2} \left[ 1 + {\cal O} \left ( \frac{1}{ \lambda N_c} \right ) \right  ]   F^1_{B B_{ X}}(q^2) \, .
\eeqa

\subsection{Numerical results}

\subsubsection{Baryon wavefunctions}

\begin{figure}[ht] 
\begin{center}
\epsfig{file=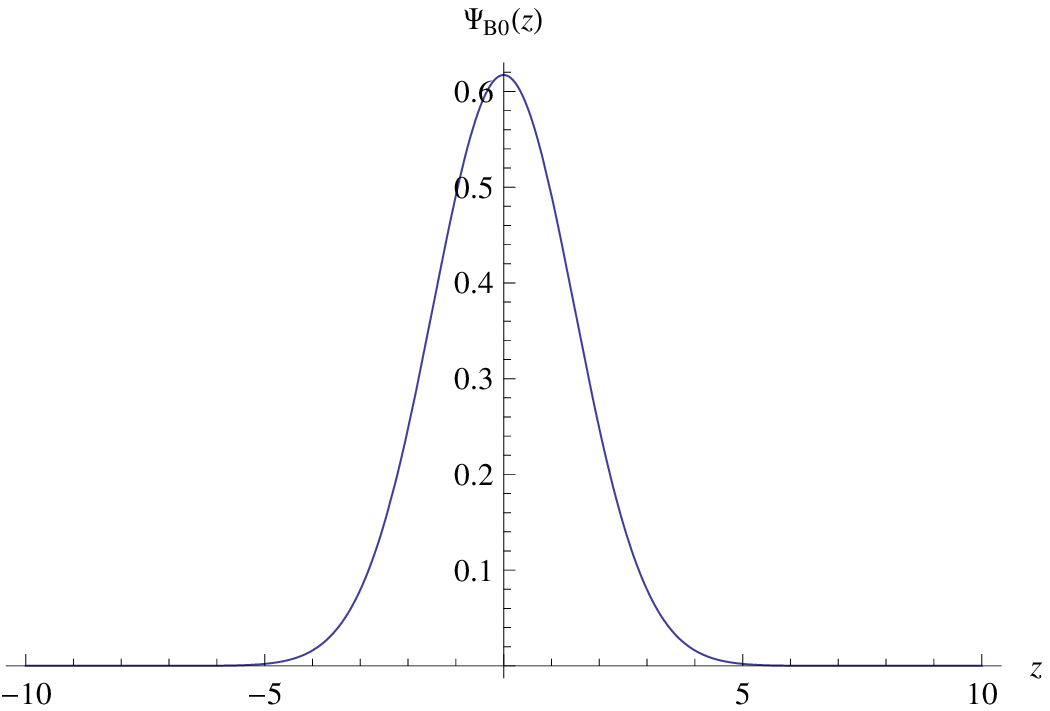, width=6cm}
\epsfig{file=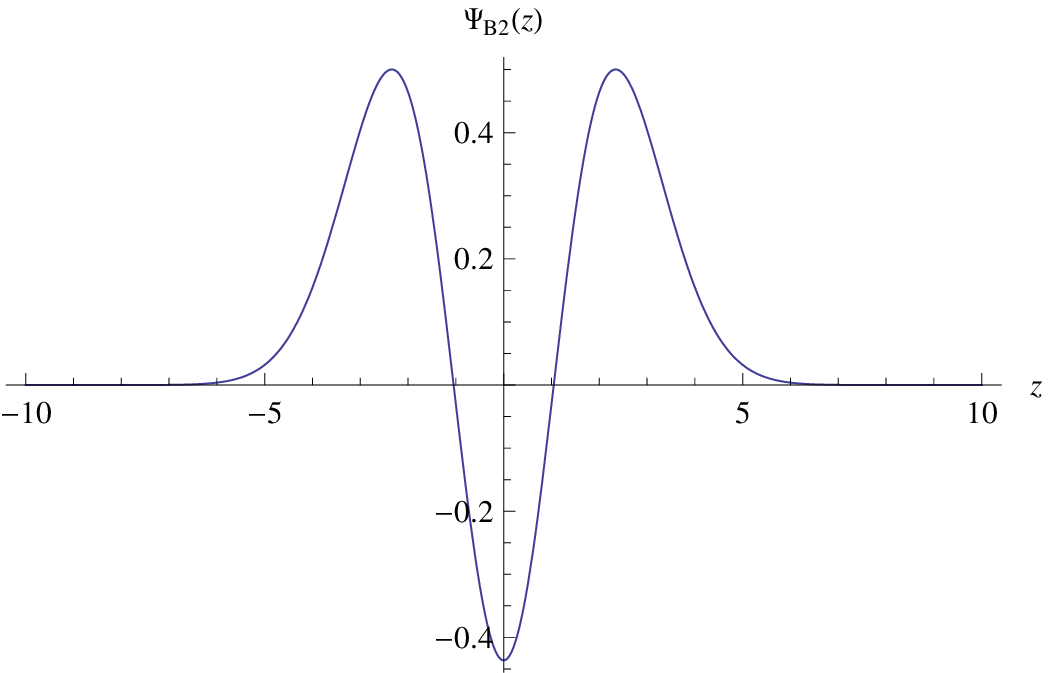, width=6cm}
\epsfig{file=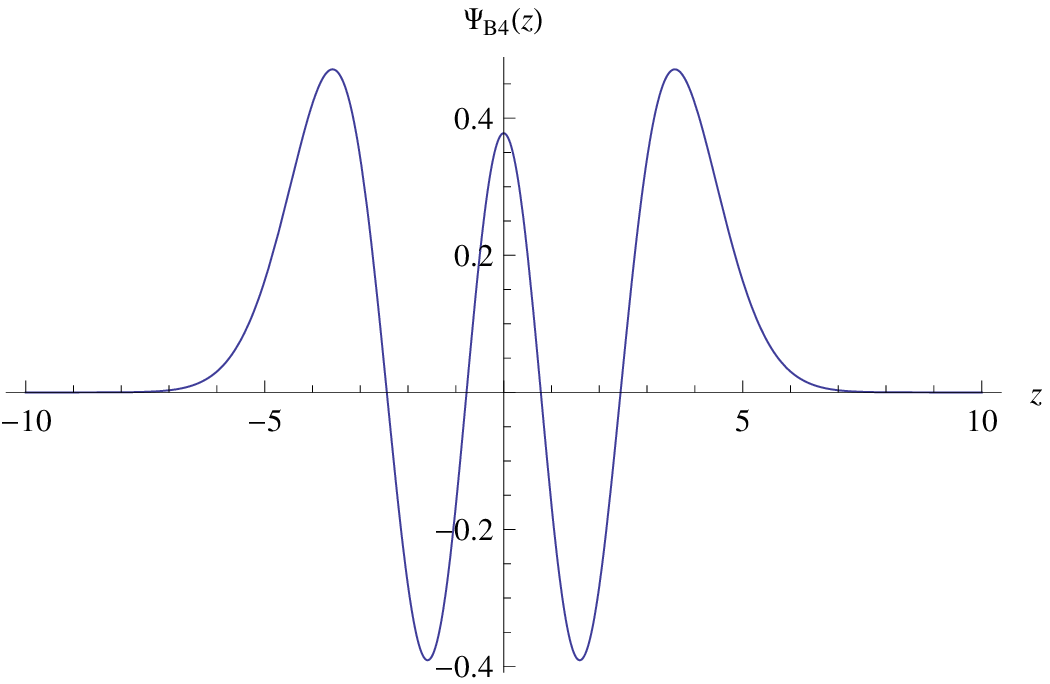, width=6cm}
\epsfig{file=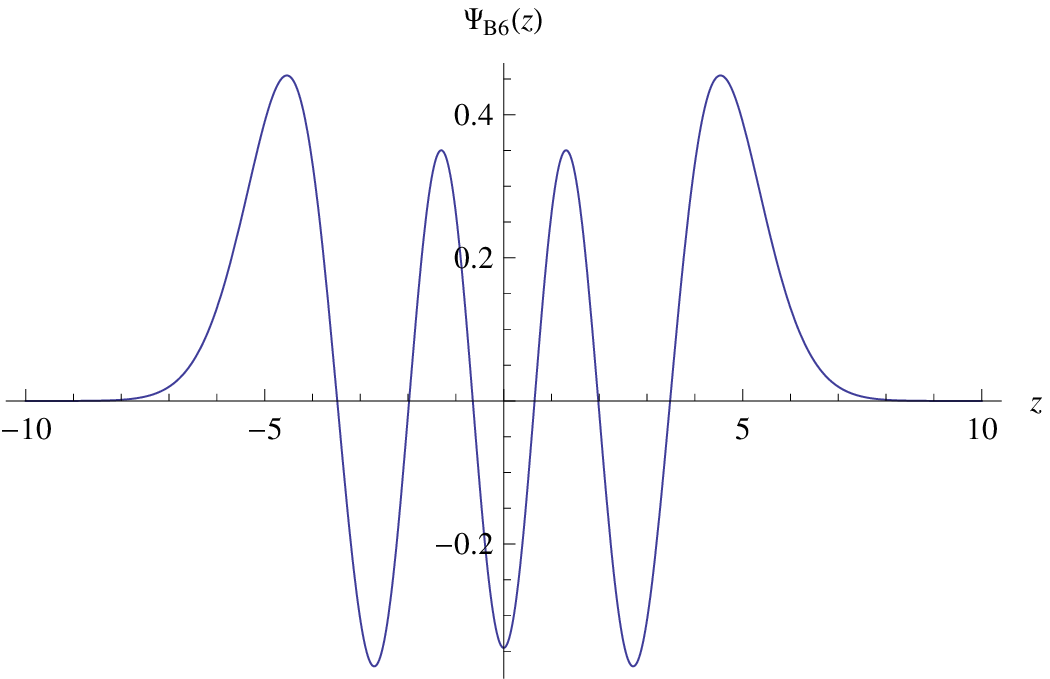, width=6cm}
\epsfig{file=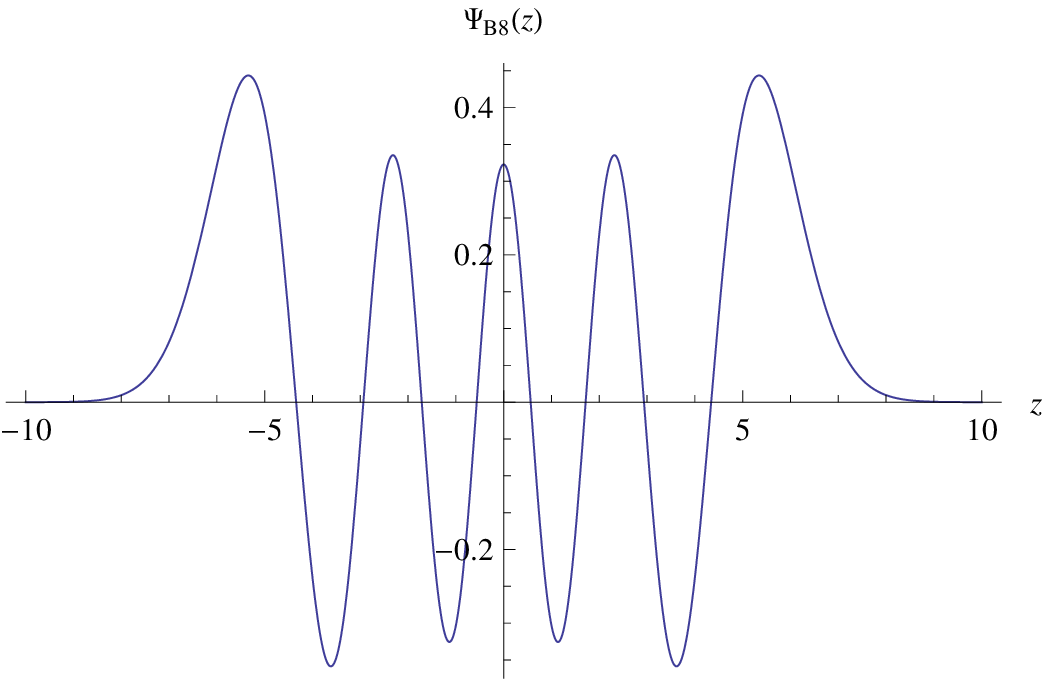, width=6cm}
\epsfig{file=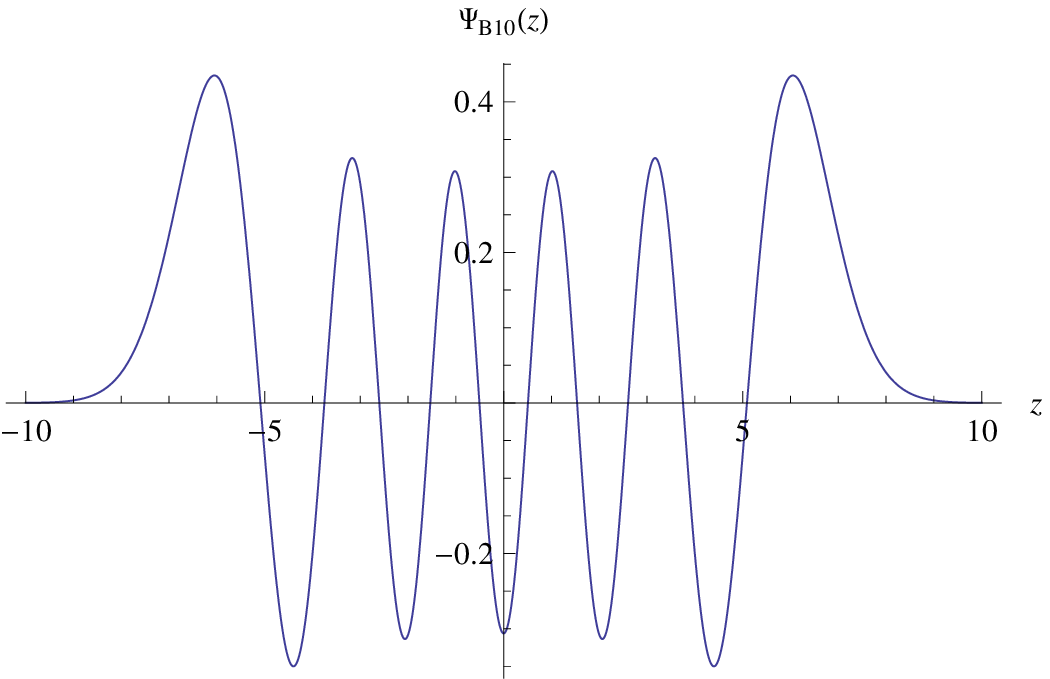, width=6cm}
\end{center}
\caption{(Normalized) wave functions $ \Psi_{B_{2k}}(z)$ for the first six even baryon states.}
\label{fig:baryonwavefct}
\end{figure}

Here, we present the results for some low-lying baryon wavefunctions in fig. \ref{fig:baryonwavefct}. With our framework, we restrict our presentation to the parity even baryon wavefunctions $n=2j$, because, as we shall see below, the coupling constants under consideration involving an even baryon state (e.g. the proton) and an odd baryon state will yield zero. 
Furthermore, we are only considering baryon states with $(n_{\rho})_{\text{initial}}=(n_{\rho})_{\text{final}}$, which is zero in the case of the proton; all other possibilities will also produce vanishing results in the calculation of the form factors below.
Table \ref{tab:baryonmasses} summarizes the mass spectrum for the proton $B=(1,+1/2,0,0)$ and its excitations $B_X=(1,+1/2,0,n_z)$. Note that we have chosen the proton mass $m_{B_0}={\wt M}_0= 940$ MeV for obvious phenomenological reasons, although strictly speaking the masses are all proportional to $N_c$ in the holographic limit $N_c, \lambda \rightarrow \infty$.\footnote{As noted in \cite{Hata:2007mb}, changing $M_{\ast}$ to a value of $500$ MeV would result in a fairly realistic mass spectrum for the excited baryon states, cf. their table (5.35). However, changing $M_{\ast}$ would alter the baryon wave functions, vector meson decay constants, coupling constants and so forth. Therefore it is not permissible to merely adapt the mass spectrum by changing $M_{\ast}$.}
\begin{table}[ht]
\begin{center}
 \begin{tabular}{|c||c|c|c|c|c|c|c|c|c||}
 \hline
 $n$ &0 & 1 & 2 & 3 & 4 & 5 & 6 & 7 & 8 \\ \hline
 $m_{B_n}$/GeV & 0.940 & 1.715 & 2.490 & 3.265 & 4.039 & 4.814 & 5.589& 6.655 &7.472 \\\hline 
\end{tabular}
\end{center}
\caption{Some numerical values for the masses of the excited baryon states connected to the proton.}\label{tab:baryonmasses}
\end{table}
  
\subsubsection{Baryon form factors}

According to figure 1, we need to study the interaction of the baryons with (external) photons. This process is represented  by the electromagnetic form factors. 
The Dirac and Pauli form factors were discussed at length in section 2. Here we present our numerical results for the generalized baryon form factors. The infinite sums over vector meson states appearing in the 
mathematical description of form factors were approximated by including the first 48 vector mesons states in the numerical computations.
The wavefunctions $\psi_{2k-1}(z)$ of the vector mesons were discussed at length in ref. \cite{BallonBayona:2009ar}.\\
Some numerical results for these quantities are listed in table \ref{tab:barCC}.
\begin{table}[ht]
\begin{center}
\begin{tabular}{|c||c|c|c|c|c|c|c|c|c||}
\hline
$k$ & 1 & 2 & 3 & 4 & 5 & 6 & 7 & 8 & 9 \\\hline\hline
$\frac{m^2_{v^k}}{M_{\ast}^2}$ & 0.6693 & 2.874 & 6.591 & 11.80 & 18.49 & 26.67 & 36.34 & 47.49 & 60.14 \\\hline\hline
$\frac{g_{v^k}}{\sqrt{\kappa}M_{\ast}^2}$ & 2.109 & 9.108 & 20.80 & 37.15 & 58.17 & 83.83 & 114.2 & 149.1& 188.7  \\\hline\hline
$g_{v^kB_0B_0}$  & {5.767} & {-2.610} & {0.1902} & {0.7664} & {-0.5162} & {-0.01955} & {0.2118} & {-0.08413} & {-0.05348} \\\hline
$g_{v^kB_0B_2}$  & {-0.9276} & {2.4670} & {-2.6239} & {1.0560} & {0.6404} & {-0.9990} & {0.2499} & {0.3848} & {-0.3049} \\\hline
$g_{v^kB_0B_4}$  & {0.3655} & {-1.2608} & {2.0708} & {-1.9409} & {0.6966} & {0.6716} & {-0.9855} & {0.2490} & {0.4558} \\\hline
$g_{v^kB_0B_6}$  & {-0.1871} & {0.7299} & {-1.4596} & {1.8556} & {-1.3815} & {0.1713} & {0.8448} & {-0.8371} & {0.03738}  \\\hline
$g_{v^kB_0B_8}$  & {0.1091} & {-0.4595} & {1.0352} & {-1.5643} & {1.5657} & {-0.7918} & {-0.3319} & {0.9318} & {-0.5634}   \\\hline\hline
\end{tabular}
\end{center}
\caption{Dimensionless squared masses and decay constants for vector mesons and coupling constants between vector mesons and baryons.}\label{tab:barCC}
\end{table}
It is interesting to compare our results with experimental data. For the vector mesons, the mass ratios for the first excited states in our framework are $m_{v^2}/m_{v^1}=2.07$ and  $m_{v^3}/m_{v^1}=3.14$. 
These results can be compared with the experimental values~\cite{Nakamura:2010zzi} $(m_{v^2}/m_{v^1})_{\text{exp}}=1.88$ and  $(m_{v^3}/m_{v^1})_{\text{exp}}=2.20$. For higher excited resonances the discrepancy between 
our results and the experimental data increases. This should be expected since the spectrum of vector mesons follows a nearly quadratic Regge trajectory in this model~\cite{BallonBayona:2009ar}. 
The coupling constant describing the elastic interaction of a $\rho$ meson with a proton can be compared with experimental data.
 In our framework we obtain $g_{v^1 B_0B_0}= 5.77$ that is consistent with experimental value $g_{v^1 B_0 B_0}^{\text{exp}} = 4.2 - 6.5$ \cite{Hohler}. The coupling constant that describes the decay of the 
resonance $P_{11}(1710)$  into a $\rho$ meson and a proton is $g_{v^1 B_0 B_2}=-0.93$ in our framework. This value is compatible with the empirical value  $|g_{v^1 B_0 B_2}^{\text{emp}}| = 0.6$ obtained from the 
experimental decay width \cite{Liu:2003hu}. There is no available experimental data for higher order couplings. 

\begin{figure}[h!] 
\begin{center}
\epsfig{file=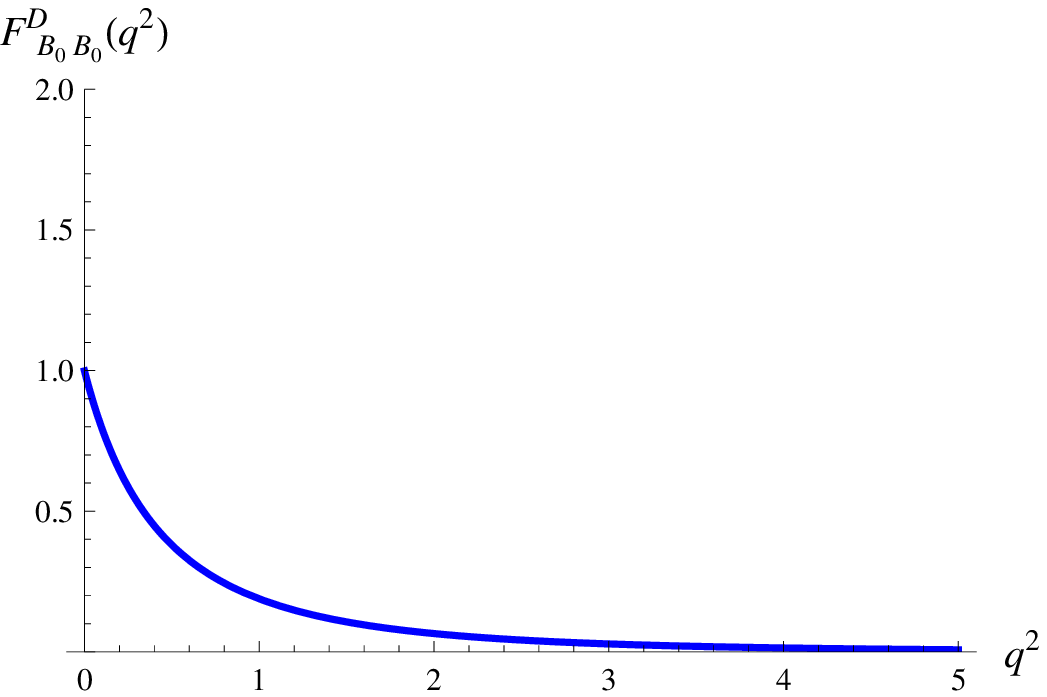, width=6cm}
\epsfig{file=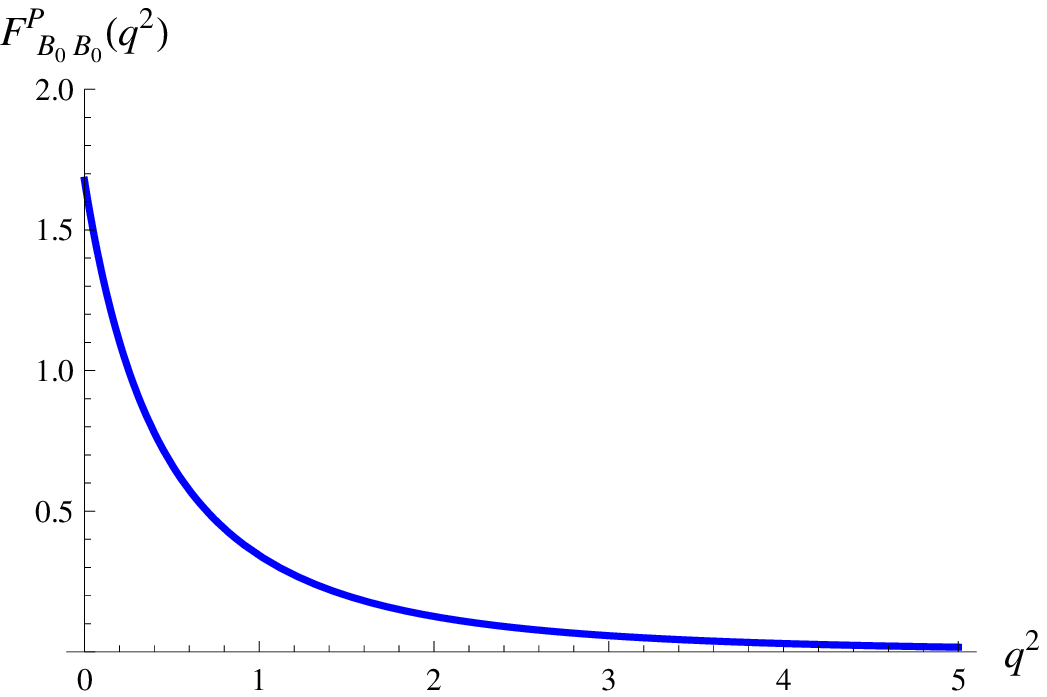, width=6cm}
\epsfig{file=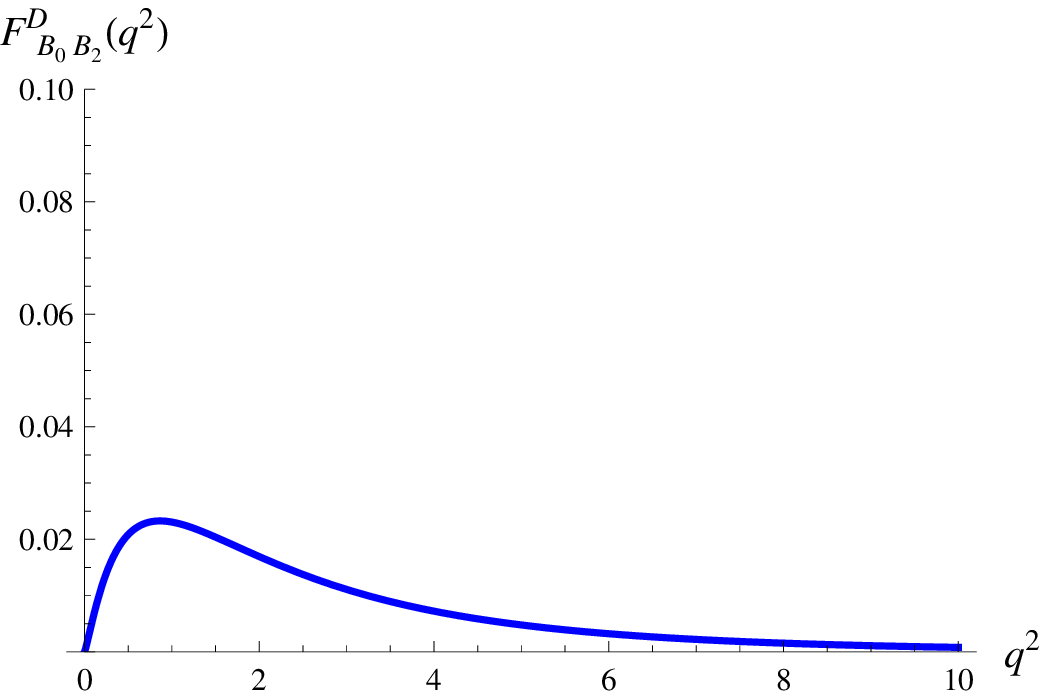, width=6cm}
\epsfig{file=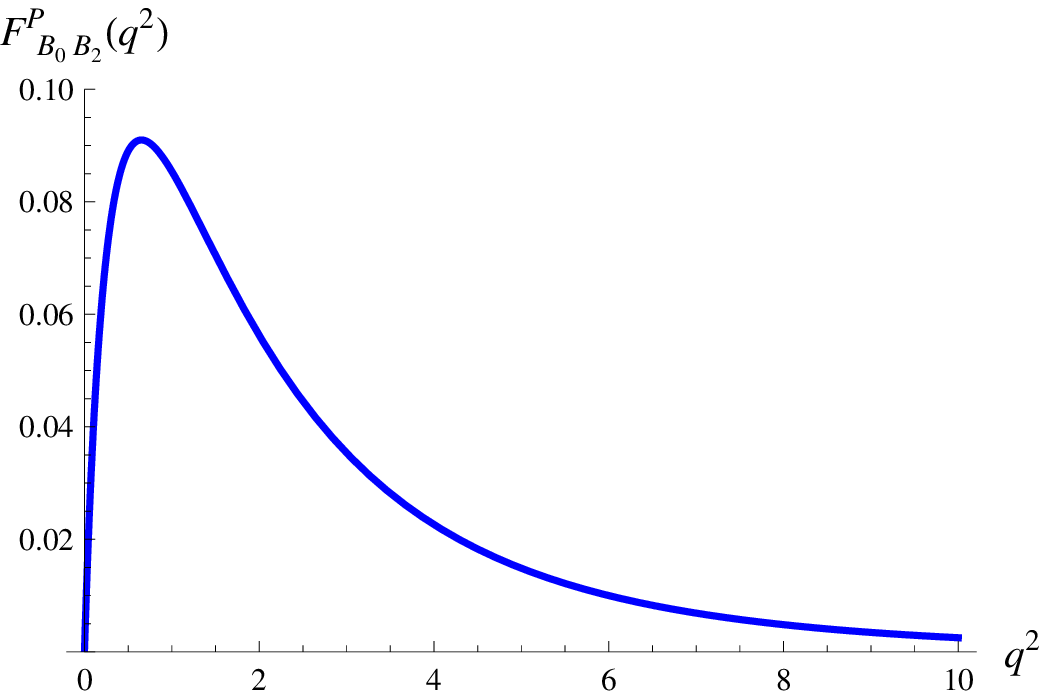, width=6cm}
\epsfig{file=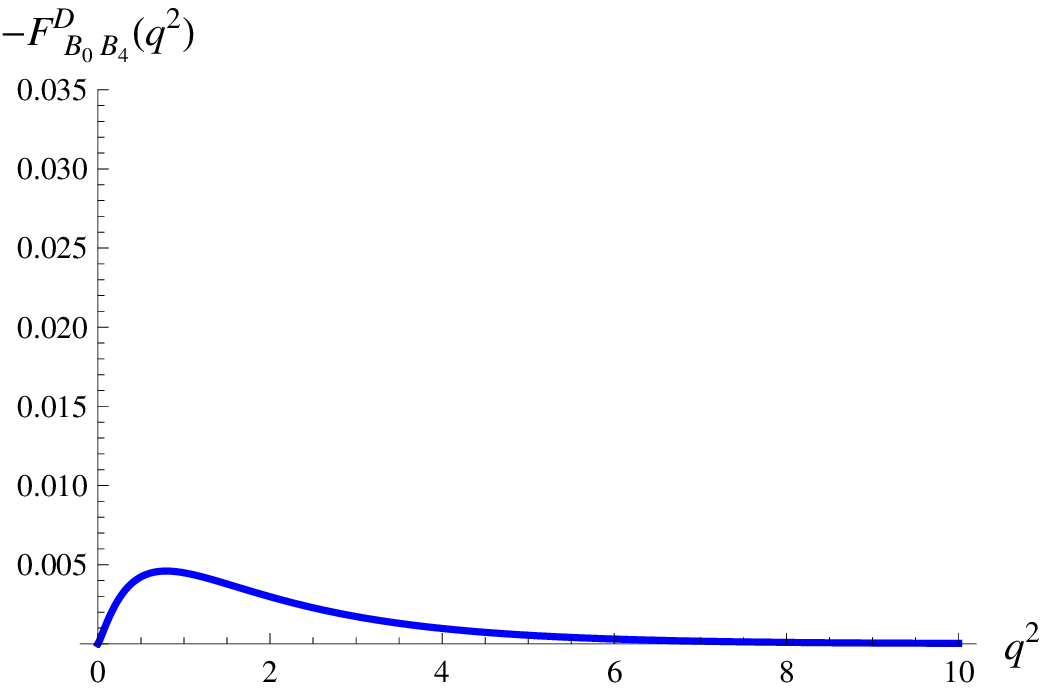, width=6cm}
\epsfig{file=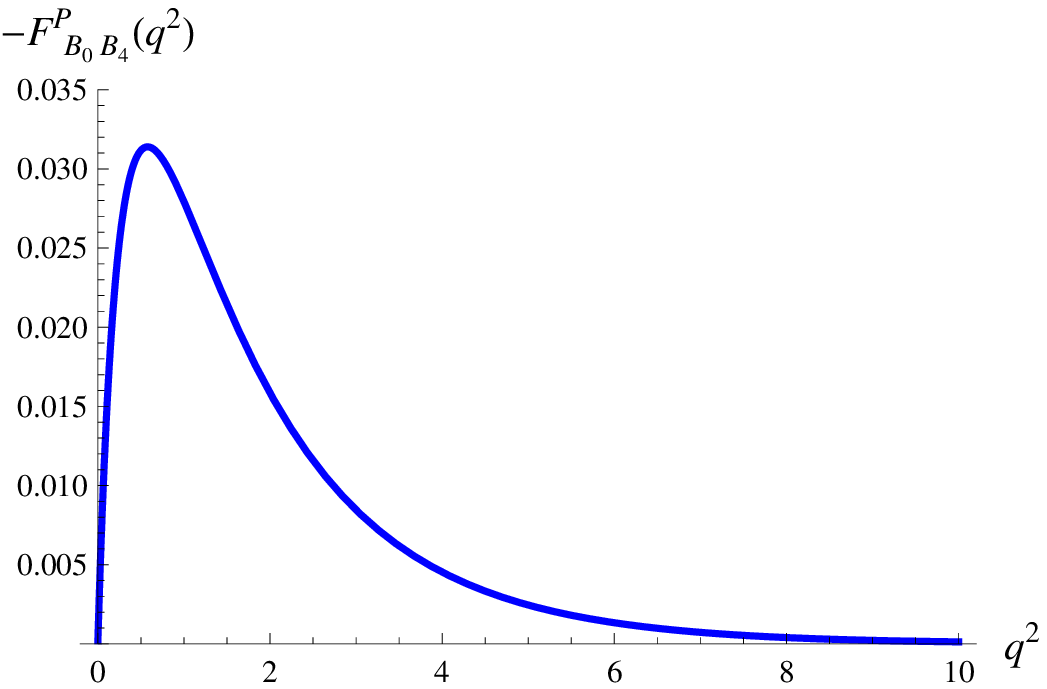, width=6cm}
\epsfig{file=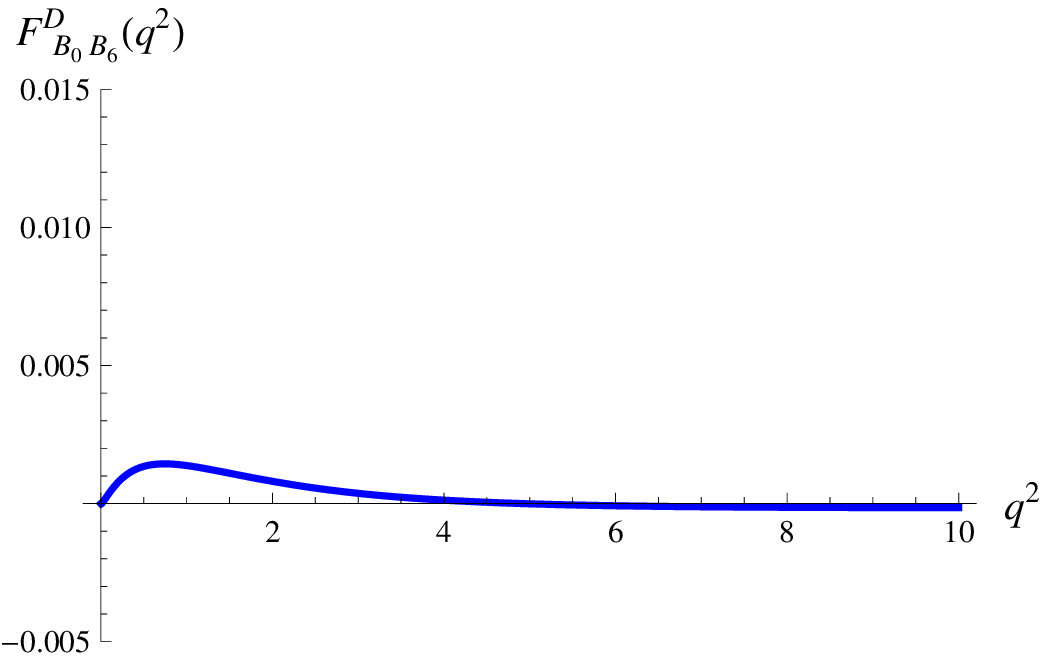, width=6cm}
\epsfig{file=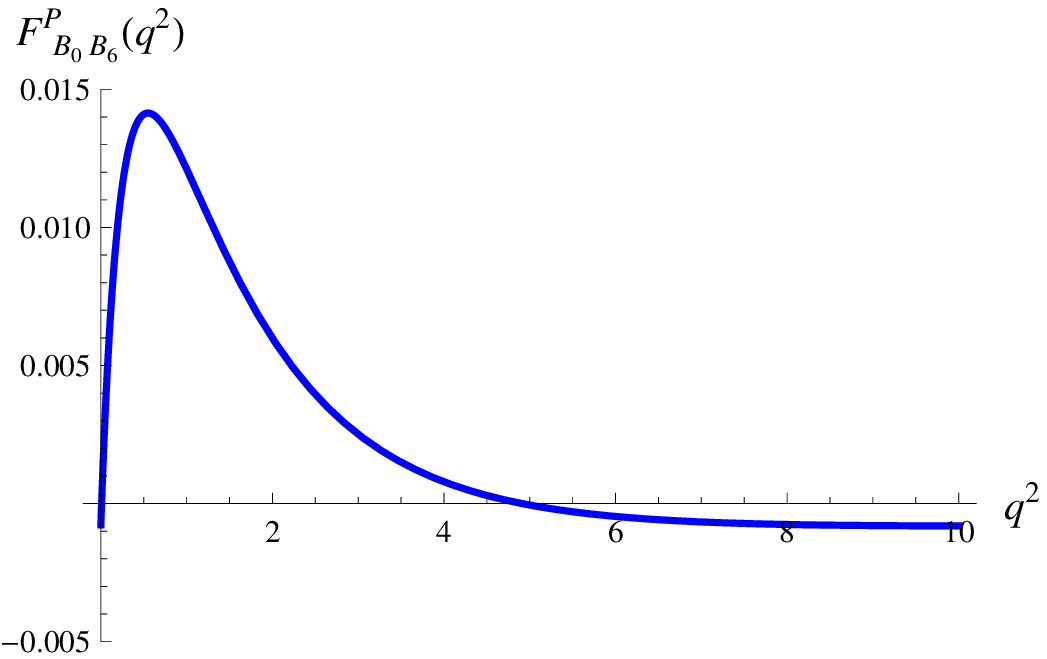, width=6cm}
\end{center}
\caption{Dirac and Pauli form factors $F^{D,P}_{B_0 B_{2j}}(q^2)$ for the first four baryon states. The elastic case corresponds to $j=0$, while $j>0$ yields the transitional form factors. For $j\geq4$, the numerical errors become relatively large; we can only trust our results for approx. $q^2 \leq 5 \,(\text{GeV})^2$.}
\label{fig:FF}
\end{figure}

Using the results for masses and coupling constants we can now easily obtain the Dirac and Pauli form factors for the first few baryon states, as shown in fig. \ref{fig:FF}. Our results for the 
elastic form factors of the proton are compatible with the results obtained in chiral soliton models~\cite{Weigel:2008zz}. Again, this should be expected from the fact that  baryons in the Sakai-Sugimoto model are
holographic representations of four dimensional skyrmions. 

Observe that the form factors $F^{D,P}_{B_0B_4}$ are negative definite. This is not problematic since they will only appear in bilinear combinations in the derivation of the structure functions below,
so that the sign will not matter.

\subsubsection{Helicity amplitudes}

\begin{figure}[h!] 
\begin{center}
\epsfig{file=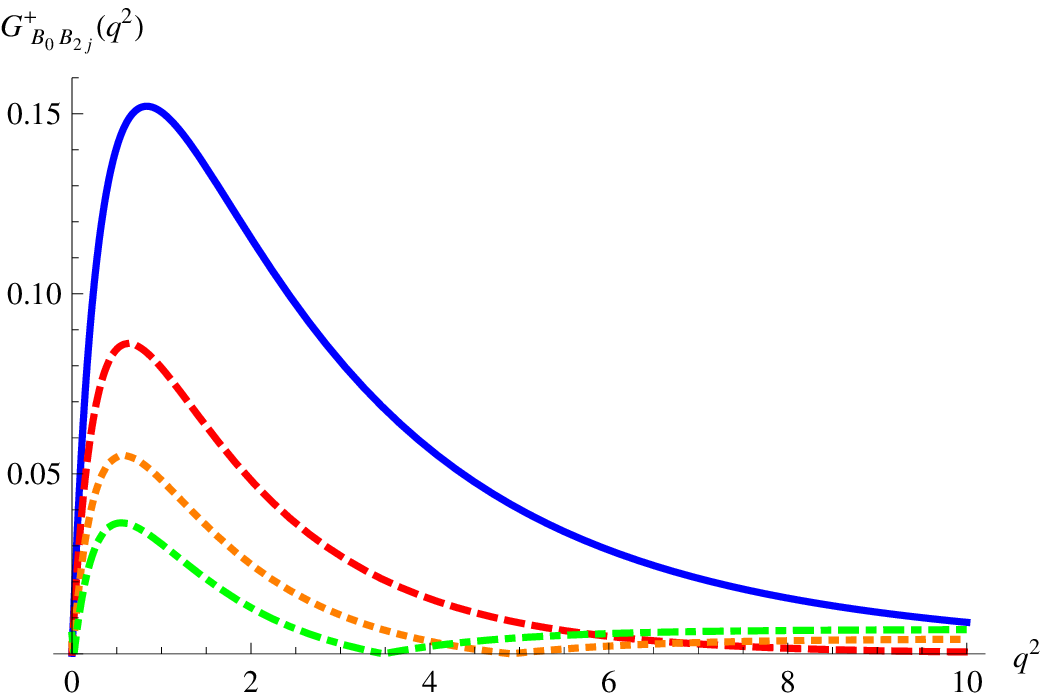, width=8cm}
\epsfig{file=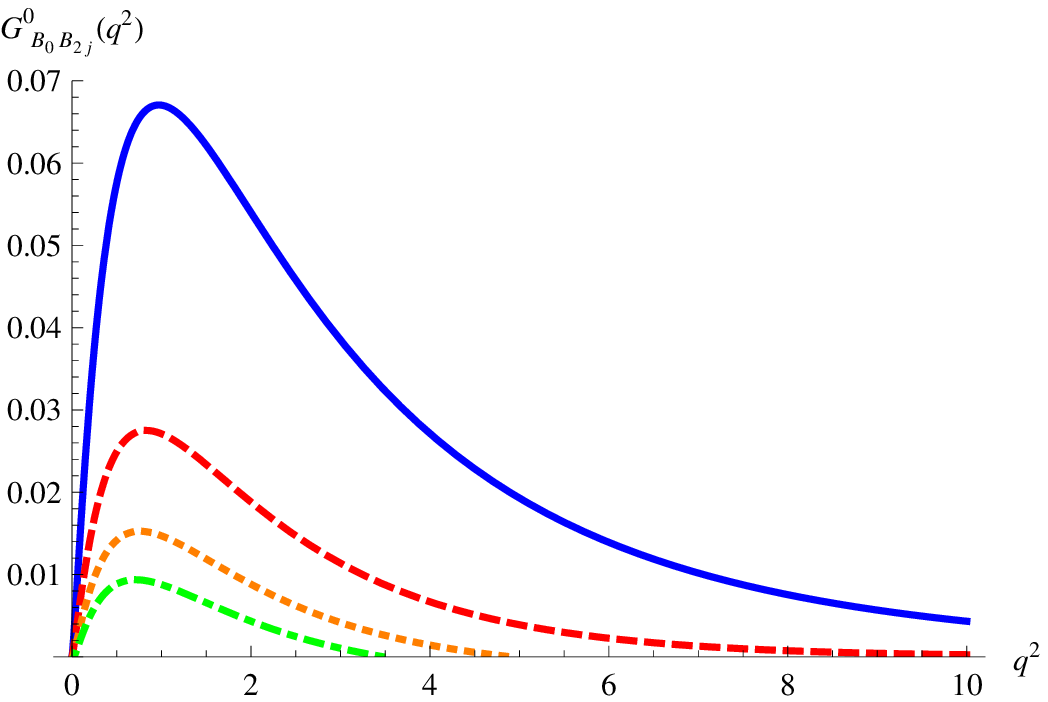, width=8cm}
\end{center}
\caption{Helicity amplitudes $|G^{+,0}_{B_0 B_{2j}}(q^2)|$ plotted versus $q^2$ in $(\text{GeV})^2$. The transitions from protons to the baryonic final states are labelled by $j$, where $j=1$ (blue, solid), $j=2$ (red, dashed), $j=3$ (orange, dotted) and $j=4$ (green, dotdashed).}
\label{fig:helamp}
\end{figure}

\begin{figure}[h!] 
\begin{center}
\epsfig{file=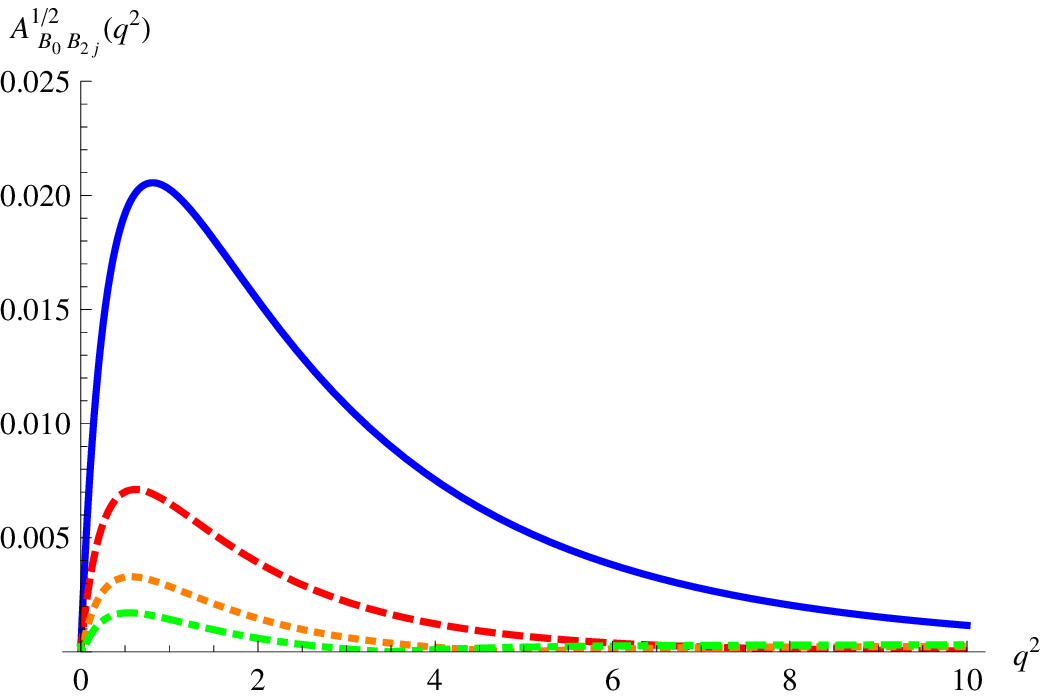, width=8cm}
\end{center}
\caption{Helicity amplitudes $|A^{1/2}_{B_0 B_{2j}}(q^2)|$ plotted versus $q^2$ in $(\text{GeV})^2$. As before, the transitions from protons to the baryonic final states are labelled by $j$, where $j=1$ (blue, solid), $j=2$ (red, dashed), $j=3$ (orange, dotted) and $j=4$ (green, dotdashed).}
\label{fig:A12}
\end{figure}

In figure \ref{fig:helamp} we present our numerical results for the helicity amplitudes $|G^+_{B B_X}(q^2)|$ and $|G^0_{B B_X}(q^2)|$ that have been discussed in section \ref{helicityamp}. 
As before, we study the helicity amplitudes in the large $\lambda$ limit, where the expressions simplify to
\beqa
(G^+_{B B_X}(q^2))^2 &=& \left [ \frac{q}{\sqrt{2} m_B} \left(1+\frac{2}{3} n_{B_X}^2 \frac{M_{\ast}^2}{q^2}\right)^{1/2} ( F^D_{B B}(q^2) + F^P_{B B}(q^2) ) \right ]^2 \, , \cr
(G^0_{B B_X}(q^2))^2 &=&   \left ( 1 + \frac{q^2}{4m_B^2 x^2}  \right ) (F^D_{B B}(q^2))^2 \,.
\eeqa
The same limitations as for the generalized form factors apply here as well, i.e., 
the numerical errors become significant for $j\geq 4$ and $q^2 \geq 5 (\text{GeV})^2$. Therefore the increase observed in $|G^+_{B B_X}(q^2)|$ for $j=3,4$ and $q^2$ greater than $4 (\text{GeV})^2$ may be an artefact 
of the numerics.
Knowledge of the helicity amplitudes $|G^{+}_{B_0 B_{2j}}(q^2)|$ is sufficient to determine the helicity amplitudes $|A^{1/2}_{B_0 B_{2j}}(q^2)|$. cf. e.g. \cite{Carlson:1998gf}.
The results are visualized in figure \ref{fig:A12}. 
The model at hand predicts $A^{1/2}_{B_0 B_{2j}}(q^2 \rightarrow 0) = 0 ({\rm GeV})^{-1/2}$, which is consistent with experimental data, cf. e.g., \cite{Nakamura:2010zzi,Aznauryan:2009mx}.
The experimental result quoted in the PDG summary tables~\cite{Nakamura:2010zzi} is
\beqa
A_{1/2}(q^2=0) = 0.009 \pm 0.022 ({\rm GeV})^{-1/2},
\eeqa 
for the transition between $P_{11}(1710)$ and the ground state. 
Other theoretical, i.e., quark model predictions are available in the literature, e.g., by Capstick and Arndt {\it et. al.} \cite{Capstick:1992uc, Arndt:1995ak},
 \beqa
A_{1/2}(q^2=0) = 0.013 ({\rm GeV})^{-1/2},
\eeqa 
for the above mentioned transition.

\section{The proton structure functions}\label{sec:PSF}

\subsection{Baryon Regge trajectories in the Sakai-Sugimoto model}

Assuming approximate continuity of the mass distribution, we can now approximate the delta distributions in the following way:
\begin{eqnarray}
\sum_{B_{\bf X}} \delta [ m_{B_{\bf X}}^2 -s ] 
&\equiv& \sum_{n} \delta [ m_{n}^2 -m_{\bar n}^2 ] 
= \int dn 
\left[\left| \frac{\partial m_n^2}{\partial n} \right|\right]^{-1}
\delta(n-\bar n)\cr
&=&\left[\left| \frac{\partial m_n^2}{\partial n} \right|\right]_{n=\bar n}^{-1}\equiv f(\bar n), \label{approxDelta} 
\end{eqnarray}
with the definition 
\beqa\label{eq:s}
s := -(p+q)^2 = m_{B_0}^2 + q^2 ( \frac 1x -1).
\eeqa
Therefore we have to evaluate the Regge trajectory of the baryon spectrum in order to calculate $\frac{\partial m_n^2}{\partial n}$. 
We find (cf. eq. (\ref{eq:masses}))
\beqa
\frac{\partial m_n^2}{\partial n} = \left(\frac{4}{\sqrt{6}}{\wt M}_0 M_{\ast}+ \frac{4}{3} n M_{\ast}^2\right), 
\eeqa
where ${\wt M}_0$ can be chosen to match, e.g. the proton mass $m_{B_0}$ and $n:= n_z$.\\

We obtain an asymptotically quadratic Regge behavior for the baryon resonances in accordance with expectations from the quantized Skyrme model,
 which is a feature inherited by the Sakai-Sugimoto model, cf. eq. (3.5). However, this is different from the experimentally observed linear behavior 
of the Regge trajectory. This is a well-known shortcoming of many  holographic models of vector mesons and baryons (see for instance 
\cite{Erdmenger:2007cm,hep-th/0610135}). There exist holographic models that exhibit a more realistic linear Regge trajectory by considering a dilaton 
field in the bulk, e.g. the soft-wall models \cite{Karch:2006pv,Gutsche:2011vb}. \\

\subsection{Numerical results}

\subsubsection{Dependence of $F_{1,2}$ on $q^2$ and $x$}
\begin{figure}[h] 
\begin{center}
\epsfig{file=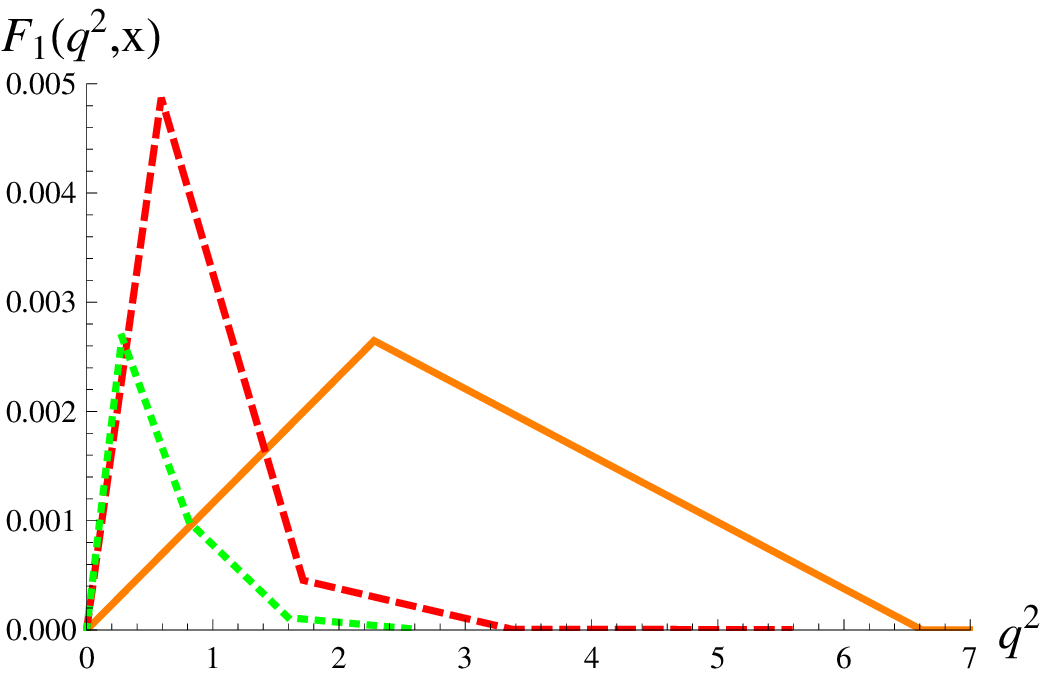, width=8cm}
\epsfig{file=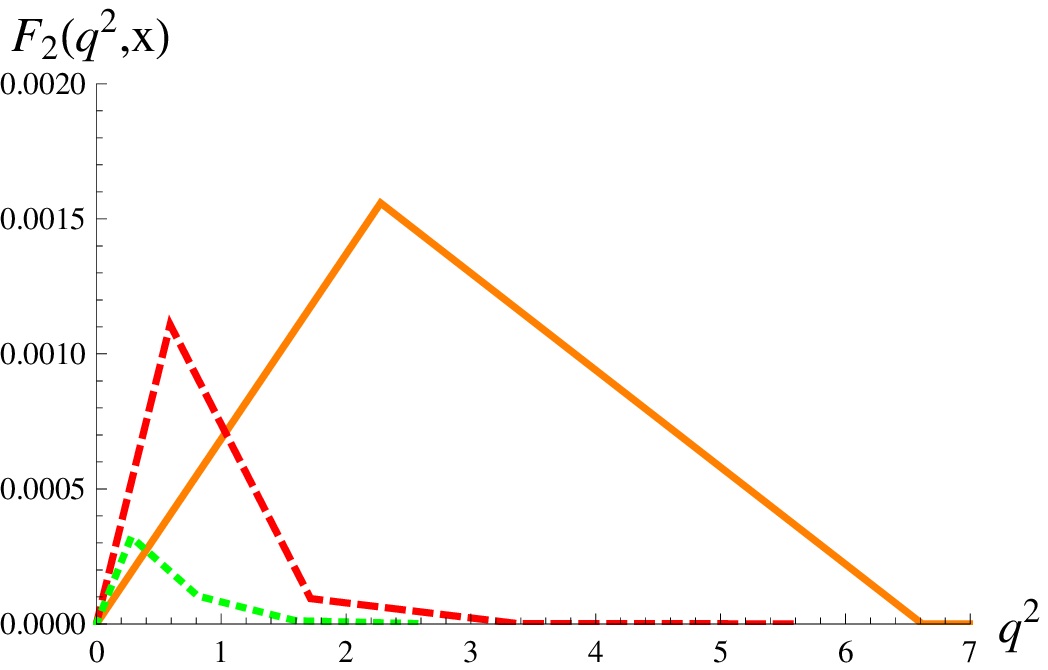, width=8cm}
\end{center}
\caption{Structure functions $F_{1,2}(q^2)$ for $x=0.3$ (orange, solid), $x=0.1$ (red, dashed) and $x=0.05$ (green, dotted).}
\label{fig:F12x}
\end{figure}

\begin{figure}[h] 
\begin{center}
\epsfig{file=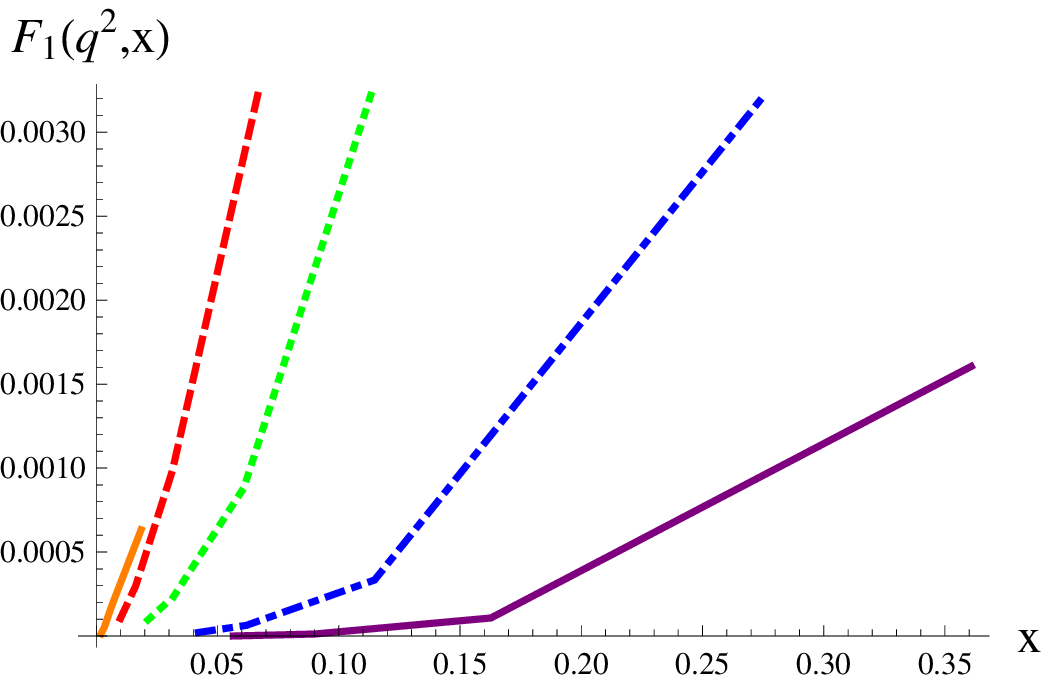, width=8cm}
\epsfig{file=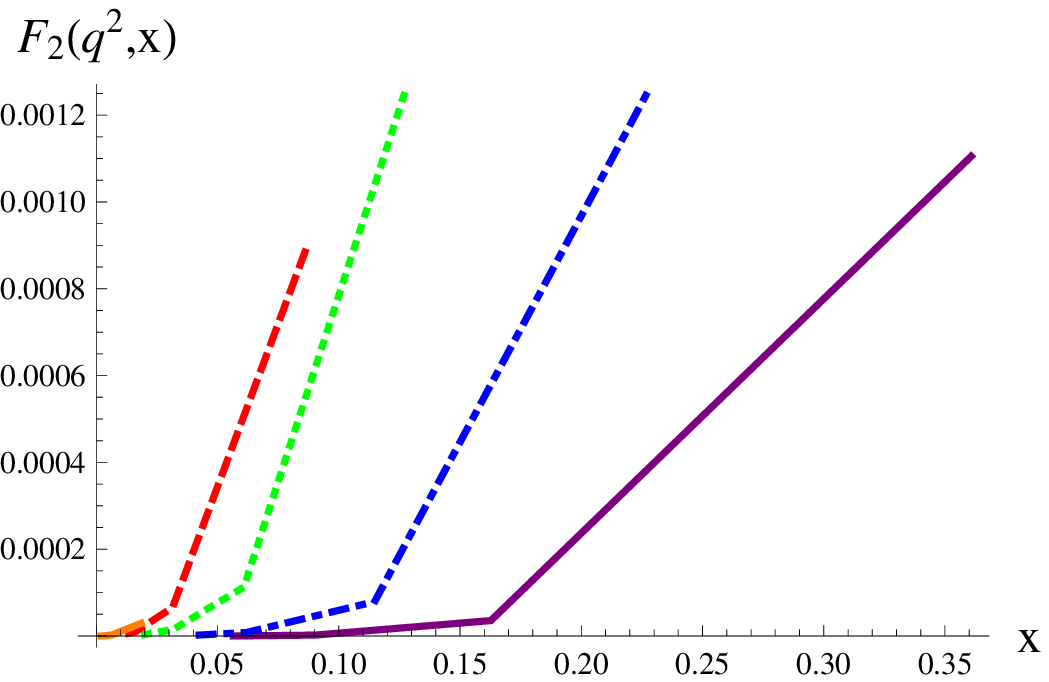, width=8cm}
\end{center}
\caption{Structure functions $F_{1,2}(x)$ for $q^2=3 (\text{GeV})^2$ (purple, solid), $q^2=2  (\text{GeV})^2$ (blue, dotdashed), $q^2=1  (\text{GeV})^2$ (green, dotted), $q^2=0.5 (\text{GeV})^2$ (red, dashed) and $q^2=0.1 (\text{GeV})^2$ (orange, solid, barely visible) .}
\label{fig:F12q}
\end{figure}

It is now possible to extract information about the structure functions $F_{1,2}(q^2, x)$ employing the following strategy: From the discrete set of baryon mass states and the relation (\ref{eq:s}), we find
\beqa\label{eq:DeltaM}
\Delta m^2_{B_{2j}}:=m^2_{B_{2j}}-m^2_{B_0}= q^2 \left( \frac{1}{x}-1\right).
\eeqa
It is possible for each $j>0$ to extract a discrete set of values for $q^2$ for a given set of fixed Bj{\o}rken parameters, e.g., $x= 0.001,\,0.01,\,0.05,\,0.1,\,0.3$. Alternatively, we may calculate a 
set of values for $x$ for a given set of, e.g., $q^2=0.1,\,0.5,\,1,\,2,\,3$ $(\text{GeV})^2$. The corresponding values are summarized in table \ref{tab:qxvalues}. 
\begin{table}[h]
\begin{center}
\begin{tabular}{|c||c|c|c|c|c||}
\hline
$j$ & $f(\bar{n}=2j)/M^2_{\ast}$&$\Delta m^2_{B_{2j}}/(\text{GeV})^2$ & $q^2/(\text{GeV})^2$ & $x$  \\\hline\hline
1 & 0.233 & 5.317 & $\begin{array}{c}  0.00532 \; \text{for}\; x=0.001 \\ 0.0537 \; \text{for}\; x=0.01 \\ 0.280 \; \text{for}\; x=0.05 \\ 0.591 \; \text{for}\; x=0.1 \\ 2.279 \; \text{for}\; x=0.3 \end{array}$ & $\begin{array}{c}  0.0185 \; \text{for}\; q^2 =0.1 (\text{GeV})^2 \\ 0.0860  \; \text{for}\; q^2 =0.5(\text{GeV})^2 \\ \; 0.158 \text{for}\; q^2 =1 (\text{GeV})^2\\ 0.273 \; \text{for}\; q^2 =2 (\text{GeV})^2 \\ 0.361 \; \text{for}\; q^2 = 3 (\text{GeV})^2   \end{array}$\\\hline
2 & 0.144 & 15.430 & $\begin{array}{c}  0.0154 \; \text{for}\; x=0.001 \\ 0.156 \; \text{for}\; x=0.01 \\ 0.812 \; \text{for}\; x=0.05 \\ 1.714 \; \text{for}\; x=0.1 \\ 6.613 \; \text{for}\; x=0.3 \end{array}$ & $\begin{array}{c}  0.00644 \; \text{for}\; q^2 =0.1 (\text{GeV})^2 \\ 0.0314  \; \text{for}\; q^2 =0.5(\text{GeV})^2 \\ \; 0.0609 \text{for}\; q^2 =1 (\text{GeV})^2\\ 0.115 \; \text{for}\; q^2 =2 (\text{GeV})^2 \\ 0.163 \; \text{for}\; q^2 = 3 (\text{GeV})^2   \end{array}$\\\hline
3 & 0.0814 & 30.353 &  $\begin{array}{c}  0.0304 \; \text{for}\; x=0.001 \\ 0.307 \; \text{for}\; x=0.01 \\ 1.598 \; \text{for}\; x=0.05 \\ 3.373 \; \text{for}\; x=0.1 \\ 13.008 \; \text{for}\; x=0.3 \end{array}$ & $\begin{array}{c}  0.00328 \; \text{for}\; q^2 =0.1 (\text{GeV})^2 \\ 0.0162  \; \text{for}\; q^2 =0.5(\text{GeV})^2 \\ \; 0.0319 \text{for}\; q^2 =1 (\text{GeV})^2\\ 0.0618 \; \text{for}\; q^2 =2 (\text{GeV})^2 \\ 0.0899 \; \text{for}\; q^2 = 3 (\text{GeV})^2   \end{array}$\\\hline
4 & 0.0436 & 54.947 & $\begin{array}{c}   0.0550 \; \text{for}\; x=0.001 \\ 0.555 \; \text{for}\; x=0.01 \\ 2.892 \; \text{for}\; x=0.05 \\ 6.105 \; \text{for}\; x=0.1 \\ 23.549 \; \text{for}\; x=0.3 \end{array}$ & $\begin{array}{c}  0.00182 \; \text{for}\; q^2 =0.1 (\text{GeV})^2 \\ 0.00902  \; \text{for}\; q^2 =0.5(\text{GeV})^2 \\ \; 0.0179 \text{for}\; q^2 =1 (\text{GeV})^2\\ 0.0351 \; \text{for}\; q^2 =2 (\text{GeV})^2 \\ 0.0518 \; \text{for}\; q^2 = 3 (\text{GeV})^2  \end{array}$\\\hline\hline
\end{tabular}
\end{center}
\caption{Some values for $q^2$ and $x$ according to eq. (\ref{eq:DeltaM}).}\label{tab:qxvalues}
\end{table}
Lastly, we need to collect some results about the calculation of the isoscalar and isovector magnetic moments for the states under consideration. For the proton and its excited states with $n_{\rho}=0$ and spin up, we find (cf. eqs. (3.16) and (3.32) of \cite{Hashimoto:2008zw})
\beqa
\mu^i_{I=0}=\frac{1}{4 M_0}\delta^{3i}\approx 0.842\, \delta^{3i} \,\mu_B, \quad \mu^i_{I=1}= \frac{M_0}{3} \frac{\sqrt{5}+ 2 \sqrt{5+N_c^2}}{2 N_c}\,\rho^2_{\text{cl}}\,\delta^{3i} \approx 3.52 \, \delta^{3i} \, \mu_B,
\eeqa
measured in units of the Bohr magneton $\mu_B = 1/(2 m_{B_0})$. Here, the mass and size of the classical instanton are $M_0 = 8 \pi^2 \kappa M_{\ast}= 558$ MeV and $\rho^2_{\text{cl}}=\frac{N_c}{8 \pi^2 \kappa M^2_{\ast}}\frac{\sqrt{6}}{\sqrt{5}}$, respectively, and we have set $N_c =3$ for obvious phenomenological reasons.
Sometimes it will be more convenient to utilize magnetic $g_I$ factors, which can be defined as follows,
\beqa
\mu^i_{I} = \frac{g_{I}}{4 m_{B_X}} \sigma^i,
\eeqa
where $\sigma^i$ are Pauli matrices. The numerical values in the Sakai-Sugimoto model turn out to be
\beqa
g_{I=0} \approx 1.684 \, , \qquad g_{I=1} \approx 7.031 \,.
\eeqa
It is now a fairly straightforward exercise to evaluate the structure functions $F_{1,2}(q^2,x)$ for the discrete set of values given in table \ref{tab:qxvalues}. Again, it should be stressed that we work 
in the large $\lambda$ limit, i.e., we only keep the leading terms in the large $\lambda$, large $N_c$ expansion {\it before} setting the baryon masses (which are of order $\mathcal{O}(\lambda N_c)$) 
to there phenomenological values.   
The results are presented in figures \ref{fig:F12x} and \ref{fig:F12q}. 

\begin{figure}[h] 
\begin{center}
\epsfig{file=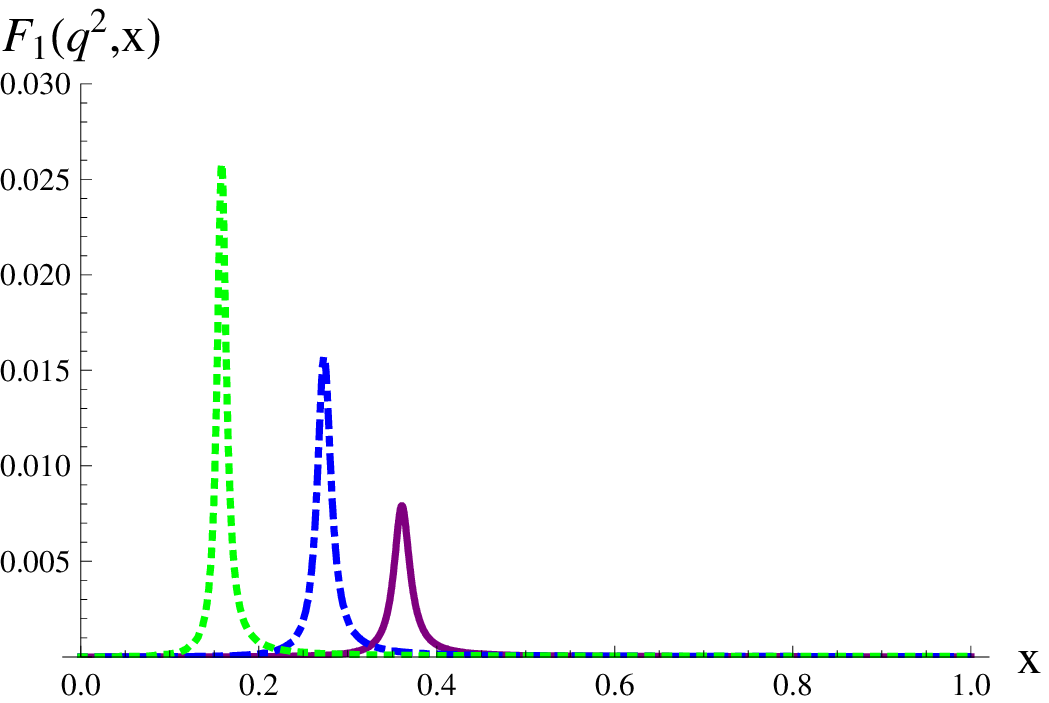, width=8cm}
\epsfig{file=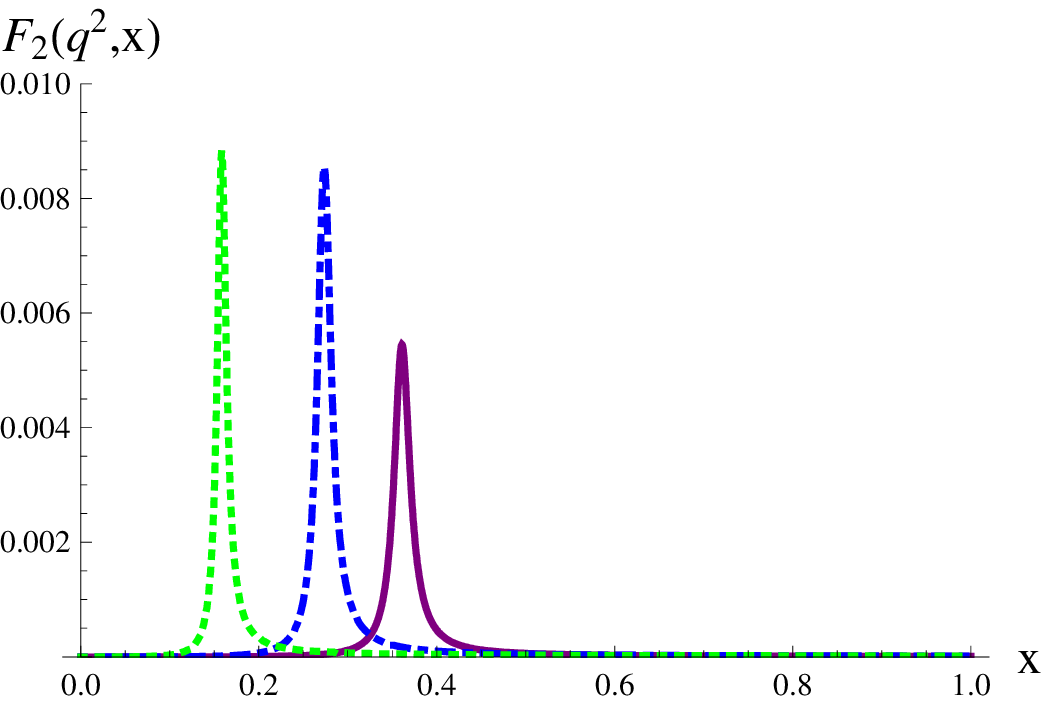, width=8cm}
\end{center}
\caption{Structure functions $F_{1,2}(x)$ for $q^2=3 (\text{GeV})^2$ (purple, solid), 
$q^2=2  (\text{GeV})^2$ (blue, dotdashed), $q^2=1  (\text{GeV})^2$ (green, dotted) near the first
 positive parity resonance $B_2$}
\label{fig:ResF1F2vsx}
\end{figure}

\begin{figure}[h] 
\begin{center}
\epsfig{file=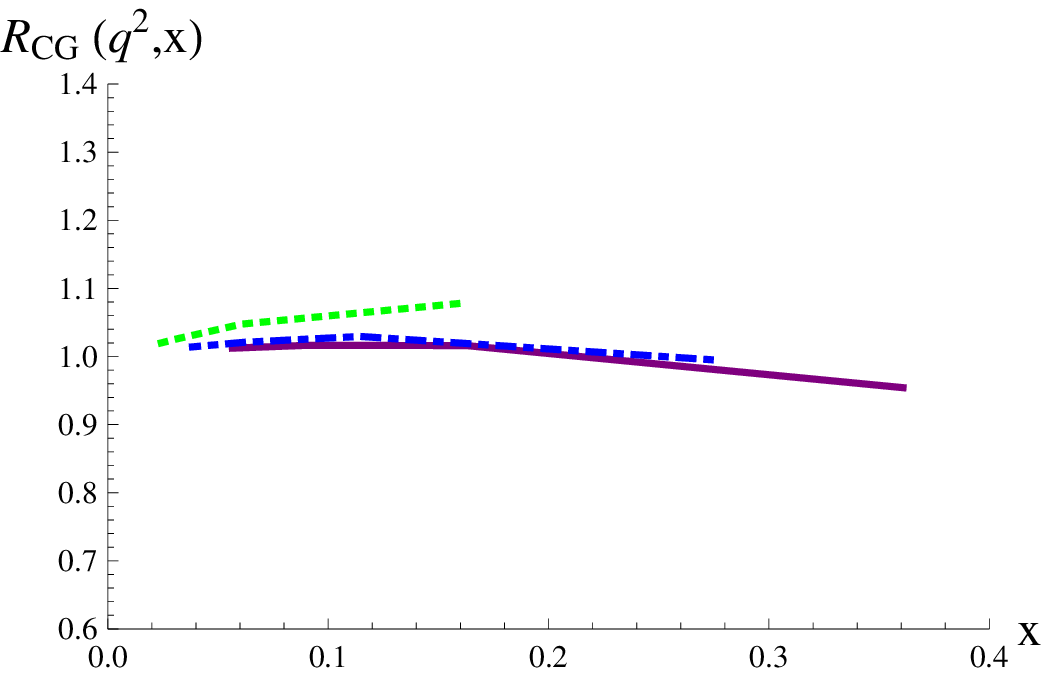, width=8cm}
\end{center}
\caption{Callan-Gross ratio $R_{\text{CG}}(x)$ for  $q^2=3 (\text{GeV})^2$ (purple, solid), $q^2=2  (\text{GeV})^2$ (blue, dotdashed) and $q^2=1  (\text{GeV})^2$ (green, dotted).}
\label{fig:CGrel}
\end{figure}

\subsubsection{Comparing our results with experimental data}

The results  shown in Figures \ref{fig:F12x} and \ref{fig:F12q} indicate that  our structure functions $F_{1,2}(q^2,x)$   seem to be at least two orders of magnitude smaller 
than what is expected from DIS experimental data (cf., e.g., \cite{Nakamura:2010zzi}, chapter 16). A possible factor that may contribute to the discrepancy between our results and experimental data 
is the naive approximation of the Delta distribution that we 
used in eq. (\ref{approxDelta}). Alternatively, we can approximate the Delta 
distribution by a Lorentzian function \cite{Carlson:1998gf} :
\beqa
\delta [ m_{B_{\bf X}}^2 -s ] \approx \frac{\Gamma_{B_{\bf X} }}{ 4 \pi m_{B_{\bf X}} } 
\left [ (\sqrt{s} - m_{B_{\bf X}})^2 + \frac{\Gamma_{B_{\bf X}}^2}{4}  \right ]^{-1} \, ,
\eeqa
where $\Gamma_{B_{\bf X}}$ is the decay width of the resonance $B_{\bf X}$. Identifying the first positive parity 
resonance $B_2$ with the experimentally observed $P_{11}(1710)$ and using the empirical value 
$\Gamma_{B_2} = 100 \, {\rm Mev}$ we obtain the  results for the structure functions shown in Figure \ref{fig:ResF1F2vsx}.
Note that the structure functions have improved by an order of magnitude. Unfortunately, we can not follow this procedure 
for the higher resonances because there are no experimental results available for the decay widths. 

The results for the proton structure functions obtained herein are understood to be non-inclusive and only represent a small fraction of possible final states, 
namely single final state baryons (the excited states of the proton) with spin $1/2$ and positive parity. If we include the contribution from final states with negative parity \cite{NegativeParity} as well as final states with higher spin\footnote{See \cite{arXiv:0904.3710} for transition to $\Delta$ resonances in the Sakai-Sugimoto model} and pion production relevant 
in this kinematical regime, we should get a better picture of the proton structure functions. 

\subsubsection{Callan-Gross relation}

The Callan-Gross relation can be easily studied numerically in this framework in order to check its validity for spin $1/2$ particles. We define the ratio 
\beqa
R_{\text{CG}}(q^2, x):= \frac{F_2(q^2,x)}{2x\, F_1(q^2,x)} \,,
\eeqa
and use the numerical results for the structure functions (Figures \ref{fig:F12x} and \ref{fig:F12q}). The result is shown in 
Figure \ref{fig:CGrel} and indicates an approximate Callan-Gross relation, i.e. $R_{\text{CG}} \approx 1$ for intermediate values of the Bjorken variable $x$. 

\medskip

\section{Conclusions and Outlook}

In the present paper we have derived a generalization of the notions of baryon electromagnetic form factors to non-elastic scattering when baryonic resonances with positive parity are produced. 
Subsequently, we have used the Sakai-Sugimoto model to estimate these form factors in the non-perturbative regime of large $N_c$ QCD. We have also estimated the contribution of these form factors to the 
proton structure functions. 

\medskip 

It is important to remark that, in this paper, we have considered the large $\lambda$ limit  in the Sakai-Sugimoto model where the description of baryons as solitons is derived from classical small instanton solutions, 
whose size is of order $\rho^2 \sim \lambda^{-1/2}$. There may exist $1/\lambda$ corrections that can be dominant at large distances as suggested in a recent analysis \cite{arXiv:0903.2662,arXiv:1109.4665}.
One should also bear in mind the limitations of describing baryons in the large $N_c$ limit \footnote{There occurs a phase transition at $N_c \approx 8$ that separates the small $N_c$ ($=3$) regime where
nuclear matter behaves like a quantum liquid from the large $N_c$ (holographic) regime where nuclear matter behaves like a crystalline solid \cite{Kaplunovsky:2010eh}}. 
Interestingly, in \cite{arXiv:0807.0316} holographic baryons were constructed for the 
(bottom-up) hard wall model as solitons in an $AdS_5$ spacetime with cut-off. These baryons have finite size and have the expected long-distance properties \cite{arXiv:0903.2662}. 
It would be interesting to investigate baryon resonance production in this model and recent holographic models such as \cite{Gursoy:2007cb, Gursoy:2007er} 
or the recent string models based on  the singular \cite{Kuperstein:2008cq,Bayona:2010bg} and deformed \cite{Dymarsky:2009cm,arXiv:1010.0993} conifold backgrounds. It would be also 
interesting to estimate the contribution of resonance production in other scattering processes like dilepton production in proton-proton scattering (see \cite{arXiv:1007.4362}). 

\medskip 

\section*{Acknowledgements}
The authors are partially supported by CAPES and CNPq (Brazilian funding agencies). C.A.B.B. is supported by the STFC Rolling Grant ST/G000433/1. 
The work of M.I. is supported by an IRCSET postdoctoral fellowship. We are grateful to S. Sugimoto for useful correspondence. 

\appendix

\section{Breit frame for inelastic scattering}
Consider the scattering between a virtual photon and a hadron  in the hadron rest frame. After two rotations we can set the spatial momentum of the photon to the $x^3$ direction, so that
\beqa
p^\mu &=& (m_B , 0 , 0 ,0 ) \cr
q^\mu &=& (q_0, 0 , 0 , q_3 ) \, ,
\eeqa
and we choose $q_3 > 0$.
The virtuality and Bjorken variable in this frame are given by
\beqa
Q^2 = q_3^2 - q_0^2   \quad , \quad x = - \frac{Q^2}{2 m_B q_0} \, .
\eeqa
Now we perform a boost in the $x^3$ direction so that
\beqa
p'^\mu &=& (\gamma m_B , 0 , 0 , - \beta \gamma m_B ) \cr
q'^\mu &=& (\gamma q_0 - \beta \gamma q_3 , 0 , 0 , - \beta \gamma q_0 + \gamma q_3 ) \,.
\eeqa
The Breit frame is defined by the condition $q'_0 =0$ so that
\beqa
\beta = \frac{q_0}{q_3}= \frac{q_0}{\sqrt{q_0^2 + Q^2}} \quad , \quad \gamma = \frac{\sqrt{q_0^2 + Q^2}}{Q} \quad , \quad q'_3 = Q  \,,
\eeqa
and we arrive at
\beqa
p'^\mu &=& (\sqrt{m_B^2 + p_3^2} , 0 , 0 , p_3 ) \cr
q'^\mu &=& (0 , 0 , 0 , Q ) \,,
\eeqa
with
\beqa
p_3 = - \frac{Q}{2x} \,.
\eeqa

\end{document}